\RequirePackage{etoolbox}
\csdef{input@path}{{style/}{graphics/}}
\documentclass[MSNbibl,nameyear,dvips]{arxstspdf}
\usepackage{graphicx}
\usepackage{url,breakurl}
\usepackage{flushend}
\usepackage{stfloats}

\volume{29}
\issue{4}
\pubyear{2014}
\firstpage{596}
\lastpage{618}
\doi{10.1214/14-STS499} % kopijuoti is laisko
\docsubty{FLA}

\makeatletter
\newcommand{\lleft}{\left}
\newcommand{\rright}{\right}

\newcommand{\bbeta}{\beta}
\newcommand{\bGamma}{\Gamma}
\newcommand{\boldO}{O}
\newcommand{\boldL}{L}
\newcommand{\boldR}{R}

\newcommand{\hatPr}{\operatorname{\hat{P}r}}
\newcommand{\ir}{\operatorname{ir}}

\newcommand{\UR}{\operatorname{UR}}
\newcommand{\Prr}{\operatorname{Pr}}
\newcommand{\NDE}{\operatorname{NDE}}
\newcommand{\cum}{\operatorname{cum}}
\newcommand{\NIE}{\operatorname{NIE}}
\newcommand{\GATE}{\operatorname{GATE}}
\newcommand{\CB}{\operatorname{CB}}

\makeatother

\begin{document}
\begin{frontmatter}

\title{Nonparametric Bounds and Sensitivity Analysis of Treatment
Effects}%\thanksref{T1}
% kai straipsnis turi susijusiu diskusiju ir rejoinder'iu
%rejoinder at \relateddoi{r}{10.1214/00-STSXXXX}.}
\runtitle{Nonparametric Bounds and Sensitivity Analysis}

\begin{aug}
\author[A]{\fnms{Amy}~\snm{Richardson}\ead[label=e1]{amyrichardson@google.com}},
\author[B]{\fnms{Michael G.}~\snm{Hudgens}\corref{}\ead[label=e2]{mhudgens@bios.unc.edu}},
\author[C]{\fnms{Peter B.}~\snm{Gilbert}\ead[label=e3]{pgilbert@scharp.org}}
\and
\author[D]{\fnms{Jason P.}~\snm{Fine}\ead[label=e4]{jfine@bios.unc.edu}}
\runauthor{Richardson, Hudgens, Gilbert and Fine}

\affiliation{Google Inc., University
of North Carolina at Chapel Hill, Fred Hutchinson Cancer Research
Center and
University of North Carolina at Chapel Hill}

\address[A]{Amy Richardson is Quantitative Analyst, Google Inc., Mountain View,
California 94043, USA \printead{e1}.}
\address[B]{Michael G. Hudgens is Associate Professor, Department of
Biostatistics, University of North Carolina at Chapel Hill, Chapel
Hill, North Carolina 27599, USA \printead{e2}.}
\address[C]{Peter B. Gilbert is Member, Statistical Center for HIV/AIDS
Research and Prevention (SCHARP),
Fred Hutchinson Cancer Research Center, Seattle, Washington 98109-1024,
USA \printead{e3}.}
\address[D]{Jason P. Fine is Professor, Department of Biostatistics,
University of North Carolina at Chapel Hill, Chapel Hill, North
Carolina 27599, USA \printead{e4}.}
\end{aug}

% ABSTRACT
%
\begin{abstract}
This paper considers conducting inference about the effect of a
treatment (or exposure) on an outcome of interest. In the ideal setting where
treatment is assigned randomly, under certain assumptions the treatment
effect is
identifiable from the observable data and inference is straightforward. However,
in other settings such as observational studies or randomized trials with
noncompliance, the treatment effect is no longer identifiable without relying
on untestable assumptions. Nonetheless, the observable data often do provide
some information about the effect of treatment, that is, the parameter
of interest
is partially identifiable. Two approaches are often employed in this
setting: (i)
bounds are derived for the treatment effect under minimal assumptions,
or (ii)
additional untestable assumptions are invoked that render the treatment effect
identifiable and then sensitivity analysis is conducted to assess how inference
about the treatment effect changes as the untestable assumptions are varied.
Approaches (i) and (ii) are considered in various settings, including assessing
principal strata effects, direct and indirect effects and effects of
time-varying
exposures. Methods for drawing formal inference about partially
identified parameters
are also discussed.
\end{abstract}

% KEYWORDS
\begin{keyword}
\kwd{Causal inference}
\kwd{nonparametric bounds}
\kwd{partially identifiable models}
\kwd{sensitivity analysis}
\end{keyword}
\end{frontmatter}

%s1 #&#
\section{Introduction}\label{sec:statscintro}

In many areas of science, interest often lies in assessing the causal
effect of
a treatment (or exposure) on some particular outcome of interest. For example,
researchers may be interested in estimating the difference between the average
outcomes when all individuals are treated (exposed) versus when all individuals
are not treated (unexposed). When treatment is assigned randomly and
there is
perfect compliance to treatment assignment, such treatment effects are
identifiable and inference about the effect of treatment proceeds in a
straightforward fashion. On the other hand, if the treatment assignment
mechanism is
not known to the analyst or compliance is not perfect, then these
treatment effects are
not identifiable from the observable data.

A statistical parameter is considered identifiable if different values
of the
parameter give rise to different probability distributions of the observable
random variables. A parameter is partially identifiable if more than
one value of the
parameter gives rise to the same observed data law, but the set of such
values is
smaller than the parameter space. Traditionally, statistical inference
has been
restricted to the situation when parameters are identifiable. More
recent research has
considered methods for conducting inference about partially identifiable
parameters. This research has been motivated to some extent by methods
to evaluate causal
effects of treatment, which are frequently partially identifiable. For
instance, causal
estimands are typically only partially identifiable in observational studies
where the treatment selection mechanism is not known to the analyst.
Noncompliance
in randomized trials may also render treatment effects partially
identifiable and a
large amount of research has been devoted to drawing inference about treatment
effects in the presence of noncompliance. Partial identifiability also arises
when drawing inference about treatment effects within principal strata
or effects
describing relationships between an outcome and a treatment that are
mediated by
some intermediate variable.

In order to conduct inference about treatment effects that are partially
identifiable, two approaches are often employed: (i) bounds are derived
for the
treatment effect under minimal assumptions, or (ii) additional untestable
assumptions are invoked under which the treatment effect is
identifiable and
then sensitivity analysis is conducted to assess how inference about the
treatment effect changes as the untestable assumptions are varied.
Below (i) and
(ii) are illustrated in five settings. In Section~\ref{sec:unmeasconfound}, we consider treatment effect
bounds and sensitivity analysis when the treatment assignment mechanism
is unknown.
In Section~\ref{sec:princstrat}, partial identifiability of principal
strata causal effects are discussed.
In Section~\ref{sec:compliance}, the setting of noncompliance is
considered where there is interest in
assessing the effect of treatment if there was perfect compliance. In
Section~\ref{sec:mediation}, bounds
and sensitivity analysis for direct and indirect effects in mediation
analysis are
presented, and in Section~\ref{sec:longtrt} longitudinal treatment
effects are considered. Much of the
literature on bounds and sensitivity analysis focuses on ignorance due
to partial
identifiability and tends to ignore uncertainty due to sampling error.
Section~\ref{sec:irur}
presents some methods that appropriately quantify this uncertainty when drawing
inference about partially identifiable treatment effects. Section~\ref{sec:statscidiscuss} concludes
with a discussion.

%%%%%%%%%%%%%%%%%%%%%%%%%%%%%%%%%%%%%%%%%%%%%%%%%%%%%%%%%%%%%%%%%%%%%%%%%%%%%%%%%%%%%%%%%%%%%%%%%%%%%%%%%%%%%%%%%%%
%%%%%%%%%%%%%%%%%%%%%%%%%%%%%%%%%%%%%%%%%%%%%%%%%%%%%%%%%%%%%%%%%%%%%%%%%%%%%%%%%%%%%%%%%%%%%%%%%%%%%%%%%%%%%%%%%%%
%%%%%%%%%%%%%%%%%%%%%%%%%%%%%%%%%%%%%%%%%%%%%%%%%%%%%%%%%%%%%%%%%%%%%%%%%%%%%%%%%%%%%%%%%%%%%%%%%%%%%%%%%%%%%%%%%%%
%%%%%%%%%%%%%%%%%%%%%%%%%%%%%%%%%%%%%%%%%%%%%%%%%%%%%%%%%%%%%%%%%%%%%%%%%%%%%%%%%%%%%%%%%%%%%%%%%%%%%%%%%%%%%%%%%%%
%%%%%%%%%%%%%%%%%%%%%%%%%%%%%%%%%%%%%%%%%%%%%%%%%%%%%%%%%%%%%%%%%%%%%%%%%%%%%%%%%%%%%%%%%%%%%%%%%%%%%%%%%%%%%%%%%%%
%%%%%%%%%%%%%%%%%%%%%%%%%%%%%%%%%%%%%%%%%%%%%%%%%%%%%%%%%%%%%%%%%%%%%%%%%%%%%%%%%%%%%%%%%%%%%%%%%%%%%%%%%%%%%%%%%%%
%s2 #&#
\section{Treatment Selection}\label{sec:unmeasconfound}

%s2.1 #&#
\subsection{Minimal Assumptions Bounds}\label{sec:noassbounds}
Suppose we have a random sample of individuals where each potentially receives
treatment or control. Unless otherwise indicated, let $Z$ indicate treatment
received where $Z=1$ denotes treatment and $Z=0$ denotes control.
Denote the observed
outcome of interest by $Y$.
In order to define a treatment effect on the outcome $Y$, we first define
potential outcomes for an individual when receiving treatment, denoted $Y(1)$,
and when receiving control, denoted $Y(0)$. Throughout this paper, we invoke
the stable unit treatment value assumption (SUTVA; \cite{Rubin1980}), that is, there is no
interference between units and there are no hidden (unrepresented)
forms of
treatment such that each individual has two potential outcomes $\{Y(0),
Y(1)\}$.
The no hidden forms of treatment guarantees that the observed outcome
is equal to the
potential outcome corresponding to the observed treatment,
namely that $Y=Y(z)$ for $Z=z$. Here, this will be referred to as  causal
consistency; for further discussion of causal consistency see
\citet{Pearl2010} and references therein. Once an individual receives
treatment $Z$, the potential outcome $Y(Z)$ is observed and the other potential
outcome (or counterfactual) $Y(1-Z)$ becomes missing. Assume that $n$
i.i.d. copies
of $(Z,Y)$ are observed and denoted by $(Z_i,Y_i)$ for $i=1,\ldots,n$.

In this section, we consider treatment effect bounds when the treatment
assignment
mechanism is unknown. Here, $Z$ can be
thought of as treatment selection by the individual or by nature,
rather than
random treatment assignment as in an experiment.\nocite{Morgan2007}
Define the average treatment effect ATE
to be $E[Y(1) - Y(0)]=E[Y(1)]-E[Y(0)]$
where $E$ denotes the expected value. The ATE can be decomposed as
%
%e1 #&#
\begin{eqnarray}
\label{eq:ace} &&\sum_{z=0}^1 E
\bigl[Y(1)|Z=z\bigr] \Prr[Z=z]
\nonumber
\\[-8pt]
\\[-8pt]
&&\quad {}-\sum_{z=0}^1 E\bigl[Y(0)|Z=z
\bigr] \Prr[Z=z].
\nonumber
\end{eqnarray}
Note $E[Y(z)|Z=z]=E[Y|Z=z]$ by  causal consistency. Thus,
from the
observed data $E[Y(z)|Z=z]$ and $\Prr[Z=z]$ are identifiable and can be
consistently estimated by their empirical counterparts. On the other
hand, the
observed data provide no information about $E[Y(z)|Z=1-z]$, such that
(\ref{eq:ace}) is only partially identifiable without additional assumptions.

Bounds on $E[Y(1) - Y(0)]$ can be obtained by entertaining the smallest
and largest
possible values for $E[Y(z)|Z=1-z]$. If $Y(1)$ and $Y(0)$ are not
bounded then
bounds on $E[Y(1) - Y(0)]$ will be completely uninformative, ranging from
$-\infty$ to $\infty$. Thus, informative bounds are only possible if
$Y(0)$ and
$Y(1)$ are bounded. Because any bounded variable can be rescaled to
take values in
the unit interval, without loss of generality assume $Y(z) \in[0,1]$ for
$z=0,1$. Then $0 \leq E[Y(z)|Z=1-z] \leq1$ and from (\ref{eq:ace}) it follows
that $E[Y(1) - Y(0)]$ is bounded below by setting $E[Y(1)|Z=0]=0$ and
$E[Y(0)|Z=1]=1$, which yields the lower bound
%
%e2 #&#
\begin{eqnarray}
\label{eq:noasslow} &&E\bigl[Y(1)|Z=1\bigr] \Prr[Z=1]
\nonumber
\\[-8pt]
\\[-8pt]
&&\quad {}- E\bigl[Y(0)|Z=0\bigr] \Prr[Z=0] - \Prr[Z=1].
\nonumber
\end{eqnarray}
Similarly, $E[Y(1) - Y(0)]$ is bounded above by setting $E[Y(1)|Z=0]=1$ and
$E[Y(0)|Z=1]=0$, which yields the upper bound
%
%e3 #&#
\begin{eqnarray}
\label{eq:noassup} &&E\bigl[Y(1)|Z=1\bigr] \Prr[Z=1]
\nonumber
\\[-8pt]
\\[-8pt]
&&\quad {}- E\bigl[Y(0)|Z=0\bigr] \Prr[Z=0] + \Prr[Z=0].
\nonumber
\end{eqnarray}
These bounds were derived independently by \citet{Robins1989} and \citet
{Manski1990}.
The lower and upper bounds (\ref{eq:noasslow}) and (\ref{eq:noassup})
are sharp
in the sense that it is not possible to derive narrower bounds without
additional assumptions. Note the interval formed by (\ref
{eq:noasslow}) and
(\ref{eq:noassup}) is contained in $[-1,1]$ and is of width 1. Thus,
the bounds
are informative in that the treatment effect is now restricted to half
of the
otherwise possible range $[-1,1]$. On the other hand, the bounds will always
contain the null value 0 corresponding to no average treatment effect.
That is,
without additional assumptions the sign of the treatment effect cannot be
determined from the observable data.

%s2.2 #&#
\subsection{Additional Assumptions}\label{sec:addassump}

The bounds (\ref{eq:noasslow})--(\ref{eq:noassup}) are sometimes
called the
``no assumptions'' or ``worst case'' bounds because no assumptions are made
about the effect of treatment in the population (\cite{lee2005}; \cite{Morgan2007}). The only assumptions made in
deriving (\ref{eq:noasslow}) and (\ref{eq:noassup}) are SUTVA and
that the
observed data constitute a random sample. If additional assumptions are invoked,
the treatment effect bounds may become tighter (i.e., narrower) or even
collapse to a
point (i.e., the treatment effect may become identifiable). Sometimes
these additional
assumptions will have implications that are testable based on the
observed data.
Should the observed data provide evidence against an assumption under
consideration,
then bounds should be computed without making this assumption.

An example of an additional assumption is mean independence, that is,
%
%e4 #&#
\begin{equation}
\label{eq:mind}\quad  E\bigl[Y(z)|Z=0\bigr] = E\bigl[Y(z)|Z=1\bigr] \quad \mbox{for }
z=0,1.
\end{equation}
Under (\ref{eq:mind}) ATE is identifiable. Specifically the upper and lower
bounds for ATE both equal
$E[Y(1)|Z=1]-E[Y(0)|Z=0]$,
which is identifiable from the observable data and can be consistently estimated
by the ``naive'' estimator given by the difference in sample means
between the groups of
individuals receiving treatment and control.
Assumption (\ref{eq:mind}) will hold in experiments where treatment is randomly
assigned as in a randomized clinical trial. Moreover, in randomized
experiments the stronger assumption
%
%e5 #&#
\begin{equation}
\label{eq:ign} Y(z) \amalg Z\quad \mbox{for } z=0,1,
\end{equation}
will hold, where $\amalg$ denotes independence. Independent treatment assignment (\ref{eq:ign}) implies mean independence~(\ref{eq:mind}).

In some settings it may be reasonable to consider additional
assumptions that
are not as strong as (\ref{eq:mind}) or (\ref{eq:ign}) but
nonetheless lead
to tighter bounds than (\ref{eq:noasslow}) and (\ref{eq:noassup}).
For example,
monotonicity type assumptions might be considered, such as monotone treatment
selection (MTS)
%
%e6 #&#
\begin{eqnarray}
\label{eq:MTSassump} \quad E\bigl[Y(z)|Z=1\bigr] \geq E\bigl[Y(z)|Z=0\bigr]\quad \mbox{for
}z=0,1.
\end{eqnarray}
MTS assumes individuals who select
treatment will on average have outcomes greater than or equal to that of
individuals who do not select treatment under the counterfactual
scenario all
individuals selected the same $z$. \citet{Manski2000}
consider MTS when examining the effect of returning to school on wages
later in life.
For this example, MTS implies individuals who
choose to return to school will have higher wages on average compared
to individuals who choose to
not return to school under the counterfactual scenario no individuals
return to school. Alternatively, one
might assume monotone treatment response (MTR)
\[
\Prr\bigl[Y(1) \geq Y(0)\bigr]=1
\]
(\cite{Manski1997}). MTR assumes that under treatment each
individual will have a response greater than or equal to that under
control. For
instance, suppose $Z=1$ if an individual elects to get the annual influenza
vaccine and $Z=0$ otherwise, and let $Y(z)=1$ if an individual subsequently
does not develop flu-like symptoms when $Z=z$, and $Y(z)=0$ otherwise.
MTR asserts
that each individual is more or as likely to not develop flu-like
symptoms if
they are vaccinated versus if they are unvaccinated. Given to date
there is no evidence
that the annual flu vaccine enhances the probability of acquiring
influenza, MTR might be
plausible for this example.

Assuming MTS or MTR can lead to narrower bounds than (\ref{eq:noasslow})
and (\ref{eq:noassup}) because they imply additional constraints on
unobserved counterfactual
expectations.
For example, assuming MTS, $E[Y(0)|Z=1]$ is bounded below by
$E[Y(0)|Z=0]$ and $E[Y(1)|Z=0]$ is bounded above by
$E[Y(1)|Z=1]$, implying the upper bound on $E[Y(1)-Y(0)]$ is
%
%e7 #&#
\begin{equation}
\label{eq:mtsup} E\bigl[Y(1)|Z=1\bigr] - E\bigl[Y(0)|Z=0\bigr],
\end{equation}
for which the naive estimator is consistent.
Under MTS, the lower bound remains (\ref{eq:noasslow}). In contrast to
the no
assumptions bounds, assuming MTS the bounds may exclude 0, specifically when
(\ref{eq:mtsup}) is negative. MTR implies
$E[Y(1)]\geq E[Y(0)]$ which in turn implies that the ATE lower
bound is 0. Under MTR, the upper bound
remains (\ref{eq:noassup}).

%s2.3 #&#
\subsection{AZT Example}\label{sec:AZT}

To illustrate the bounds above, consider a hypothetical study of 2000
HIV patients
(from Figure~2 of \cite{Robins1989}) where 1400 individuals elected
to take the drug
AZT and 600 elected not to take AZT (this is a simplified version of
the problem
Robins considers). The outcome of interest is death or survival at a
given time
point. Of the 2000 patients, 1000 died with exactly 500 from each
group. Let
$Z=1$ if the patient elected to take AZT and $Z=0$ otherwise; let $Y=1$
if the
individual died and 0
otherwise. The naive estimator, that is, the difference in sample means between
$Z=1$ and $Z=0$, equals
$500/1400-500/600 \approx -0.48$. The empirical estimates of the no assumptions
bounds (\ref{eq:noasslow}) and (\ref{eq:noassup}) equal $-0.7$ and
$0.3$. In this setting, the
MTS assumption (\ref{eq:MTSassump}) supposes that individuals who
elected to take AZT
would have been more or as likely to die as individuals who did not
take AZT in the
counterfactual scenarios where everyone receives treatment or everyone
does not
receive treatment. This might be reasonable if it is thought that those
who took
AZT were on average less healthy than those who did not. Assuming MTS,
the upper
bound (\ref{eq:mtsup}) is estimated to be $-$0.48. Thus, in this example
the MTS
bounds are substantially tighter than the no assumption bounds. The estimated
MTS bounds lead to the conclusion (ignoring sampling variability, a
point which
we return to later) that AZT reduces the probability of death by at
least 0.48 whereas
without the MTS assumption we cannot even conclude whether the effect of
treatment is nonzero.

%s2.4 #&#
\subsection{Sensitivity Analysis}\label{sec:unmeasconfoundsensanal}

Assumptions such as (\ref{eq:mind}) or (\ref{eq:ign}) which identify
the ATE,
or assumptions such as MTS which sharpen the bounds, cannot be tested
empirically
because such assumptions pertain to the counterfactual distribution of $Y(z)$
given $Z=1-z$. Robins and others (e.g., see \cite{Robinsetal1999}; \cite{Scharfstein1999}) have argued that a data analyst
should conduct sensitivity analysis to explore how inference varies as
a function
of departures from any untestable assumptions.

For instance, a departure from assumption (\ref{eq:ign}) might be due to
the existence of an unmeasured variable $U$ associated with both treatment
selection $Z$ and the potential outcomes $Y(z)$ for $z=0,1$;
a variable such as $U$ is often referred to as an
unmeasured confounder. Under this scenario, one might postulate that
$Y(z)\amalg Z|U$ for $z=0,1$ rather than (\ref{eq:ign}). Sensitivity analysis
proceeds by examining how inference drawn about ATE
varies as a function of the magnitude of the association of $U$ with
$Z$, $Y(0)$,
and $Y(1)$. This idea has roots as early as \citet{Cornfield1959}, who
demonstrated the plausibility of a causal effect of cigarette smoking
($Z$) on
lung cancer ($Y$) by arguing that the absence of such a relationship
was only
possible if there existed an unmeasured factor $U$ associated with
cigarette use
that was at least as strongly associated with lung cancer as cigarette use.
This idea was further developed by \citet{Schlesselman1978},
\citet{Rosenbaum1983a},
\citet{Lin1998}, \citet{Hernan1999} and \citet{VanderWeele2011a} among others.

To illustrate this approach, suppose in the AZT example above that the analyst
first assumes (\ref{eq:ign}) holds, and thus estimates the effect of
AZT to be
$-$0.48. To proceed with sensitivity analysis, the analyst posits the existence
of an unmeasured binary variable $U$ and assumes that $Y(z)\amalg Z|U$ for
$z=0,1$. Similar to \citet{VanderWeele2011a},
let
\begin{eqnarray*}
c(z)&=&\bigl\{E\bigl[Y(z)|U=1\bigr]
-E\bigl[Y(z)|U=0\bigr]\bigr\}
\\
&&{}\cdot\bigl\{\Prr[U=1|Z=z]-\Prr[U=1]\bigr\}.
\end{eqnarray*}
Then under the assumption that $Y(z)\amalg Z|U$ for $z=0,1$, the naive estimator
converges in probability to $E[Y(1)] - E[Y(0)] + c(1)-c(0)$.
Thus the naive estimator is asymptotically unbiased if and only if
$c(1)=c(0)$. For an
alternative decomposition of the asymptotic bias of the naive
estimator, see
\citeauthor{Morgan2007} (\citeyear{Morgan2007}, Section~2.6.3).

Sensitivity analysis proceeds by making varying assumptions about the
unidentifiable associations of $U$ with $Y(0)$, $Y(1)$ and $Z$. Under
the most
extreme of these assumptions, the bounds (\ref{eq:noasslow}) and
(\ref{eq:noassup}) are recovered. In particular, the upper bound in
(\ref{eq:noassup}) is achieved when $\Prr[U=1|Z=1]=0$, $\Prr[U=1|Z=0]=1$,
$E[Y(1)|U=1]=1$ and $E[Y(0)|U=0]=0$, meaning that the confounder $U$ is
perfectly negatively
correlated with treatment $Z$ and that if the confounder is present
($U=1$), then a treated
individual will die, whereas if the confounder is absent ($U=0$), then
an untreated
individual will survive. The lower bound (\ref{eq:noasslow}) is
achieved under the
opposite conditions.

In practice the extreme associations of $U$ with $Y(0)$, $Y(1)$, and
$Z$ leading
to the bounds might be considered unrealistic. Instead the analyst might
consider associations only in a range deemed plausible by subject
matter experts. In order
to arrive at an accurate range, care should be taken in communicating
the meaning of
these associations and eliciting this range should be done in a manner
that avoids data
driven choices. Alternatively, the degree of associations required to
change the sign of
the effect of interest might be determined. For instance, suppose the
analyst further assumes that
$E[Y(z)|U=1] - E[Y(z)|U=0]$ does not depend on $z$. This assumption
will hold if the effect of
$Z$ on $Y$ is the same if $U=0$ or $U=1$.
Letting $\gamma_0= E[Y(z)|U=1] - E[Y(z)|U=0]$ and $\gamma_1
= \Prr[U=1|Z=1] -
\Prr[U=1|Z=0]$, the asymptotic bias of the naive estimator is then
given by $\gamma_0\gamma_1$ and a bias adjusted
estimator is found by subtracting $\gamma_0\gamma_1$ from the naive
estimator. Sensitivity analysis may
proceed by determining the values of $\gamma_0$ and $\gamma_1$ for
which the bias
adjusted estimator of the ATE will have the opposite sign of the naive
estimator.
For the AZT example, the bias adjusted estimator will have the
opposite sign of the naive estimator if $\gamma_0\gamma_1 < -0.48$. This
indicates that the product of (i) the difference in the mean potential
outcomes between
levels of the confounder for both treatment and control, and (ii) the
difference in the prevalence of the
unmeasured confounder between the treatment and
control groups must be less than $-$0.48. Such magnitudes might be
considered unlikely
in the opinion of subject matter experts, in which case the sensitivity
analysis would
support the existence of a beneficial effect of AZT on survival among
HIV+ men (ignoring
sampling variability). Note the observed data distribution places some
restrictions on the possible values of $(\gamma_0,\gamma_1)$,
that is, $(\gamma_0,\gamma_1)$ is partially identifiable. For
instance, if $\gamma_1=1$ then
$\Prr[U=1|Z=1]=1$ and $\Prr[U=1|Z=0]=0$ which implies $E[Y(z)|U=u]=
E[Y(z)|Z=u]$
and, therefore, $\max\{E[Y(1)|Z=1]-1, -E[Y(0)|Z=0]\} \leq\gamma_0
\leq\min\{E[Y(1)|Z=1], 1-E[Y(0)|Z=0]\}$. Such considerations
should be taken into account when determining the range of values of
$(\gamma_0$, $\gamma_1)$ in sensitivity analysis.

Because the data provide no evidence about $U$,
{\spaceskip=0.2em plus 0.05em minus 0.05em \citet{VanderWeele2008a}
and \citet{VanderWeele2011a}} recommend choosing
$U$ and any simplifying assumptions based on what is considered
plausible by relevant
subject-matter experts. Such sensitivity analyses are
most applicable when the existence of unmeasured confounders is known,
but these factors
could not be measured for logistical or other reasons. General bias
formulas to be used for sensitivity analyses of
unmeasured confounding for categorical or continuous outcomes,
confounders and
treatments can be found in \citet{VanderWeele2011a}.

In other settings, there might not be any known unmeasured confounders,
or it may be
thought that there are numerous unmeasured confounders, in which cases
the sensitivity
analysis strategy described above would not be applicable or feasible.
One general alternative approach entails
making additional untestable assumptions regarding the unobserved
potential outcome distributions. Typically, these
assumptions (or models)
are indexed by one or more sensitivity analysis parameters conditional
upon which the causal estimand of interest
is identifiable
(e.g., \cite{Scharfstein1999}; \cite{Brumback2004}).
Sensitivity
analysis then proceeds by examining how inference changes as assumed
values of the
parameters are varied over plausible ranges. Examples of such
sensitivity analyses are
given below in Sections~\ref{sec:pesensanal} and \ref{sec:msmsensanal}.

%s2.5 #&#
\subsection{Covariate Adjustment}\label{sec:covadjust}
Typically in observational studies baseline (pre-treatment) covariates
$X$ will be collected in
addition to $Z$ and $Y$. Incorporating information from observed
covariates can help
sharpen inferences about partially identified treatment effects. For example,
incorporating covariates will generally lead to narrower bounds
(\cite{Scharfstein1999}).
This follows because any treatment effect compatible with
the distribution of observed variables $(X,Y,Z)$ must also be
compatible with the
distribution of $(Y,Z)$, that is, the observable variables if we do not
observe or choose to
ignore $X$ (\cite{Lee2009}). Covariate adjusted bounds are discussed
further in Section~\ref{sec:pebounds} below.

Additionally, incorporating covariates may lend plausibility to some of
the bounding
assumptions discussed in Section~\ref{sec:addassump}. For example, in
the absence of randomized
treatment assignment (\ref{eq:mind}) or (\ref{eq:ign}) may be
dubious. Instead of
(\ref{eq:mind}), it might be more plausible to assume
%
%e8 #&#
\begin{eqnarray}
\label{eq:mindcov} &&E\bigl[Y(z)|Z=0,X=x\bigr]
\nonumber
\\[-8pt]
\\[-8pt]
&&\quad = E\bigl[Y(z)|Z=1,X=x\bigr]\quad \mbox{for } z=0,1.
\nonumber
\end{eqnarray}
Similarly, assumption
(\ref{eq:ign}) might be replaced by
%
%e9 #&#
\begin{equation}
\label{eq:igncov} Y(z)\amalg Z|X\quad \mbox{for } z=0,1,
\end{equation}
that is, each potential outcome is independent of treatment selection
conditional on some set of covariates. Assumption (\ref{eq:igncov}) is commonly
referred to as no unmeasured confounders. Assumptions
such as (\ref{eq:mindcov}) or weaker inequalities similar to (\ref
{eq:MTSassump}) such as
\begin{eqnarray*}
&& E\bigl[Y(z)|Z=1,X=x\bigr]
\\
&&\quad \geq E\bigl[Y(z)|Z=0,X=x\bigr] \quad \mbox{for } z=0,1,
\end{eqnarray*}
may be deemed plausible for certain levels of $X$, but not for others.
Availability of covariates also allows for
the consideration of new types of assumptions
(e.g., see \cite{Chiburis2010}).

To conduct covariate adjusted sensitivity analysis, departures from
identifying assumptions such as
(\ref{eq:igncov}) can be explored.
Similar to the previous section, a departure from (\ref{eq:igncov})
might entail positing the
existence of an unmeasured variable $U$ associated
with both treatment selection $Z$ and the potential outcomes $Y(z)$ for
$z=0,1$. Under
this scenario, one might postulate that $Y(z)\amalg Z|\{X,U\}$ for
$z=0,1$ rather than
(\ref{eq:igncov}) and sensitivity analysis proceeds by examining how
inference varies as a function of the magnitude of
the association of $U$ with $Z$, $Y(0)$, and $Y(1)$ given $X$. Similar
to covariate adjusted bounds,
smaller associations or tighter regions of the values of the
sensitivity parameters
may be deemed plausible within certain levels of $X$, potentially
yielding sharper inferences from
the sensitivity analyses. However, as cautioned by \citet{Robins2002},
care should be taken in clearly
communicating the meaning of such sensitivity parameters and their
relationship to
covariates when eliciting plausible ranges from subject matter experts.
In some scenarios, plausible regions for
sensitivity parameters may in fact be wider when conditioning on $X$
than when not conditioning on $X$.

%%%%%%%%%%%%%%%%%%%%%%%%%%%%%%%%%%%%%%%%%%%%%%%%%%%%%%%%%%%%%%%%%%%%%%%%%%%%%%%%%%%%%%%%%%%%%%%%%%%%%%%%%%%%%%%%%%%
%%%%%%%%%%%%%%%%%%%%%%%%%%%%%%%%%%%%%%%%%%%%%%%%%%%%%%%%%%%%%%%%%%%%%%%%%%%%%%%%%%%%%%%%%%%%%%%%%%%%%%%%%%%%%%%%%%%
%%%%%%%%%%%%%%%%%%%%%%%%%%%%%%%%%%%%%%%%%%%%%%%%%%%%%%%%%%%%%%%%%%%%%%%%%%%%%%%%%%%%%%%%%%%%%%%%%%%%%%%%%%%%%%%%%%%
%%%%%%%%%%%%%%%%%%%%%%%%%%%%%%%%%%%%%%%%%%%%%%%%%%%%%%%%%%%%%%%%%%%%%%%%%%%%%%%%%%%%%%%%%%%%%%%%%%%%%%%%%%%%%%%%%%%
%%%%%%%%%%%%%%%%%%%%%%%%%%%%%%%%%%%%%%%%%%%%%%%%%%%%%%%%%%%%%%%%%%%%%%%%%%%%%%%%%%%%%%%%%%%%%%%%%%%%%%%%%%%%%%%%%%%
%%%%%%%%%%%%%%%%%%%%%%%%%%%%%%%%%%%%%%%%%%%%%%%%%%%%%%%%%%%%%%%%%%%%%%%%%%%%%%%%%%%%%%%%%%%%%%%%%%%%%%%%%%%%%%%%%%%
%s3 #&#
\section{Principal Stratification}\label{sec:princstrat}

%s3.1 #&#
\subsection{Background}\label{sec:backps}

Even if treatment is randomly assigned (e.g., as in a clinical trial), the
causal estimand of interest may still be only partially identifiable. For
example, in many studies it is often of interest to draw inference about
treatment effects on outcomes that only exist or are meaningful after
the occurrence of some observable
intermediate variable. For instance, in studies where some individuals
die, investigators might be interested in
treatment effects only among individuals alive at the end of the study.
Unfortunately,
estimands defined by contrasting mean outcomes under treatment and
control that
simply condition on this observable intermediate variable do not
measure a causal effect of treatment without additional assumptions.
One approach that may
be employed in this scenario entails principal stratification
(\cite{Frangakis2002}).
Principal stratification uses the potential
outcomes of the intermediate post-randomization variable to define
strata of
individuals. Because these ``principal strata'' are not affected by treatment
assignment, treatment effect estimands defined within principal strata
have a
causal interpretation and do not suffer from the complications of standard
post-randomization adjusted estimands. The simple framework of principal
stratification has a wide range of applications. For a recent
discussion of the
utility (and lack thereof) of principal
stratification, see \citet{Pearl2011} and corresponding reader reactions.

As a motivating example for this section, we consider evaluating
vaccine effects
on post-infection outcomes. In vaccine studies, uninfected subjects are \mbox{enrolled}
and followed for infection endpoints, and infected subjects are subsequently
followed for post-infection outcomes such as disease severity or death
due to
infection with the pathogen targeted by the vaccine; often interest is in
assessing the effect of vaccination on these post-infection endpoints
(\cite{Hudgens2006}).
For example, \citet{Preziosi2003} present data from a pertussis vaccine
field study in
Niakhar, Senegal. In this study, 3845 vaccinated children and 1020 unvaccinated
children were followed for one year for pertussis. In the vaccine
group, 548
children contracted pertussis, of whom 176 had severe infections; in the
unvaccinated group 206 children contracted pertussis, of whom 129 had severe
infections. In this setting, investigators are interested in assessing
whether or
not the vaccine had an effect on the severity of infection.

When assessing such post-infection effects, a data analyst might consider
contrasts between study arms including all individuals under study, or,
alternatively, only those who become infected. Though including all
individuals in
the study has the advantage of providing valid inference about the overall
effect of vaccination (assuming independent treatment assignment), such an approach
does not
distinguish vaccine effects on susceptibility to infection from effects
on the
post-infection endpoint of interest. An analysis that conditions on infection
attempts to distinguish these effects and may be more sensitive in detecting
post-infection vaccine effects. However, because the set of individuals who
would become infected under control are not likely to be the same as
those who
would become infected if given the vaccine, conditioning on infection might
result in selection bias. For example, those who would become infected under
vaccine may tend to have weaker immune systems than those who would become
infected under control, and thus may be more susceptible to severe infection.
Because of this potential selection bias, comparisons between infected vaccinees
and infected controls do not necessarily have causal interpretations.

%s3.2 #&#
\subsection{Principal Effects}\label{sec:pe}

In this section, treatment is vaccination, with $Z=1$ corresponding to
vaccination
and $Z=0$ corresponding to not being vaccinated. Assume that assignment to
vaccine is equivalent to receipt of vaccine, that is, there is no noncompliance.
Denote the potential infection outcome by $S(z)$, where $S(z)=0$ if uninfected
and $S(z)=1$ if infected. Here, the focus is on evaluating the causal
effect of
vaccine on $Y$, a post-infection outcome. For simplicity, we consider
the case
where $Y$ is binary, indicating the presence of severe disease. If $S(z)=1$,
define the potential post-infection outcome $Y(z) = 1$ if the
individual would
have the worse (or more severe) post-infection outcome of interest
given $z$,
and $Y(z) = 0$ otherwise. If an individual's potential infection
outcome for
treatment $z$ is uninfected [i.e., $S(z)=0$], then we adopt the
convention that
$Y(z)$ is undefined. In other words, it does not make sense
to define the severity of an infection in an individual who is not
infected. This
convention is similar to that employed in other settings. For instance,
in the
analysis of quality of life studies it might be assumed that quality of life
metrics are not well defined in those who are not alive
(\cite{Rubin2000}).

Define a \textit{basic principal stratification} $P_0$ according to
the joint
potential infection outcomes $S^{P_0} = (S(0),S(1))$. The four basic principal
strata or response types are defined by the joint potential
infection outcomes, $(S(0),S(1))$, and are composed of immune (not infected
under both vaccine and placebo), harmed (infected under vaccine but not
placebo),
protected (infected under placebo but not vaccine), and doomed individuals
(infected under both vaccine and placebo). Note the only stratum where both
potential post-infection endpoints are well defined is in the doomed basic
principal stratum, $S^{P_0}=(1,1)$. Thus, defining a post-infection causal
vaccine effect is only possible in the doomed principal stratum $S^{P_0}=(1,1)$.
Such a causal estimand will describe the effect of vaccination on disease
severity in individuals who would become infected whether vaccinated or not.
For instance, the vaccine effect on disease severity may be defined by
%
%e10 #&#
\begin{eqnarray}
\label{eq:ACEy1} &&E\bigl[Y(1)|S^{P_0}=(1,1)\bigr] \nonumber\\[-8pt]\\[-8pt]
&&\quad - E\bigl[Y(0)|S^{P_0}=(1,1)
\bigr].\nonumber
\end{eqnarray}
Frangakis and Rubin call treatment effect estimands such as (\ref{eq:ACEy1})
``principal effects.''

%s3.3 #&#
\subsection{Bounds}\label{sec:pebounds}

Assume we observe $n$ i.i.d. copies of $(Z,S,Y)$ denoted by
$(Z_i,S_i,Y_i)$ for
$i=1,\ldots,n$. Also assume that the doomed principal strata is
nonempty, $\Prr[S^{P_0}=(1,1)]>0$, so that the principal effect in
(\ref{eq:ACEy1}) is well defined.
Bounds for (\ref{eq:ACEy1}) are presented below under two additional
assumptions: independent treatment assignment, that is,
%
%e11 #&#
\begin{equation}
\label{eq:ignore} Z \amalg\bigl\{Y(z),S(z)\bigr\}\quad \mbox{for } z=0,1
\end{equation}
and monotone treatment response with
respect to $S$, that is,
%
%e12 #&#
\begin{equation}
\label{eq:monos} \Prr\bigl[S(0) \geq S(1)\bigr] = 1.
\end{equation}
Assumption (\ref{eq:ignore}) will hold in randomized vaccine trials.
Monotonicity (\ref{eq:monos}) assumes that the vaccine does no harm at the
individual level, that is, there are no individuals who would be
infected if
vaccinated but uninfected if not vaccinated. Monotonicity is equivalent to
assuming the harmed principal stratum is empty. Note no such
monotonicity assumption is being made regarding $Y$. Under (\ref
{eq:ignore}), assumption
(\ref{eq:monos}) implies $P(S=1|Z=1) \leq P(S=1|Z=0)$, which is
testable using
the observed data. For the pertussis example, the proportion infected
in the vaccine group
was less than in the unvaccinated group; thus, assuming (\ref
{eq:ignore}), the data do not provide
evidence against (\ref{eq:monos}).

Assuming independent treatment assignment and monotonicity, (\ref
{eq:ACEy1}) is
partially identifiable from the observable data. The left term of (\ref
{eq:ACEy1})
can be written
%
%e13 #&#
\begin{eqnarray}
\label{eq:id} %
&&E\bigl[Y(1)|S^{P_0}=(1,1)\bigr] \nonumber\\
&&\quad = E
\bigl[Y(1)|S(1)=1\bigr]
\nonumber\\[-8pt]\\[-8pt]
&&\quad = E\bigl[Y(1)|S(1)=1,Z=1\bigr]\nonumber
\\
&&\quad = E[Y|S=1,Z=1],
\nonumber
\end{eqnarray}
where the first equality holds under (\ref{eq:monos}), the second
equality under (\ref{eq:ignore}), and the third
by causal consistency. On the other hand, the right term of
(\ref{eq:ACEy1}) is only partially identifiable. To see this, note
%
%e14 #&#
\begin{eqnarray}
\label{eq:id1} %
&&E\bigl[Y(0)|S(0)=1\bigr]
\nonumber
\\
&&\quad\hspace*{6pt} =E\bigl[Y(0)|S^{P_0}=(1,1)\bigr]
\Prr\bigl[S(1)=1|S(0)=1\bigr]
\\
&&\hspace*{24pt} {}+E\bigl[Y(0)|S^{P_0}=(1,0)\bigr] \Prr\bigl[S(1)=0|S(0)=1
\bigr].
\nonumber
\end{eqnarray}
In (\ref{eq:id1}), only $E[Y(0)|S(0)=1]$ and $\Prr[S(1)=s|\allowbreak S(0)=1]$ for
$s=0,1$ are identifiable.
In particular, $E[Y(0)|S(0)=1] = E[Y|S=1,Z=0]$ by similar reasoning to
(\ref{eq:id}), and
\begin{eqnarray*}
&&\Prr\bigl[S(1)=1|S(0)=1\bigr]
\\
&&\quad = \frac{\Prr[S(1)=1]}{\Prr[S(0)=1]}= \frac{\Prr[S=1|Z=1]}{\Prr[S=1|Z=0]},
\end{eqnarray*}
where the first equality holds under (\ref{eq:monos}) and the second under
independent treatment assignment (and causal consistency). The other
two terms
in (\ref{eq:id1}), namely $E[Y(0)|S^{P_0}=(1,1)]$ and $E[Y(0)|S^{P_0}=(1,0)]$,
are only partially identifiable. In words, infected controls are a
mixture of
individuals in the protected and doomed principal stratum and without further
assumptions the observed data do not identify exactly which infected controls
are doomed. Therefore, the probability of severe disease when not
vaccinated in
the doomed principal stratum is not identified. Under (\ref
{eq:monos}), the data
do however indicate what proportion of infected controls are doomed and this
information provides partial identification of $E[Y(0)|S^{P_0}=(1,1)]$,
and hence
(\ref{eq:ACEy1}).

For fixed values of $E[Y(0)|S(0)=1]$ and $\Prr[S(1)=1|S(0)=1]$, any
pair of
expectations $(E[Y(0)|S^{P_0}=(1,1)],E[Y(0)| S^{P_0}=(1,0)]) \in[0,1]^2$
satisfying (\ref{eq:id1}) will give rise to the same observed data
distribution.
Equation (\ref{eq:id1}) describes a line segment with nonpositive slope
intersecting the unit square as illustrated in Figure~\ref{fig:statsci1}. An upper bound of
$E[Y(0)|S^{P_0}=(1,1)]$ and thus a lower bound for (\ref{eq:ACEy1}) is achieved
when the line intersects the right or lower side of the square, that
is, when either
%
%e15 #&#
\begin{eqnarray}
\label{eq:lowerc1} E\bigl[Y(0)|S^{P_0}=(1,1)\bigr] &=&1\quad \mbox{or}
\nonumber
\\[-8pt]
\\[-8pt]
E\bigl[Y(0)|S^{P_0}=(1,0)\bigr]&=&0.
\nonumber
\end{eqnarray}
Together (\ref{eq:id1}) and (\ref{eq:lowerc1}) imply
$E[Y(0)|S^{P_0}=(1,1)]$ is
bounded above by
%
%e16 #&#
\begin{equation}
\label{eq:upper} \min \biggl\{1, \frac{E[Y(0)|S(0)=1]}{\Prr[S(1)=1|S(0)=1]} \biggr\}.
\end{equation}
Similarly,
$E[Y(0)|S^{P_0}=(1,1)]$
is bounded below by
%
%e17 #&#
\begin{eqnarray}
\label{eq:lower}
&&\max \biggl\{ 0, \nonumber\\[-8pt]\\[-8pt]
&&\hspace*{25pt}\frac{E[Y(0)|S(0)=1] - \Prr[S(1)=0|S(0)=1]}{
\Prr[S(1)=1|S(0)=1]
} \biggr\}.\nonumber
\end{eqnarray}
Combining (\ref{eq:lower}) with (\ref{eq:id}) yields the upper bound
on the
principal effect of interest (\ref{eq:ACEy1}) and combining (\ref{eq:upper})
with (\ref{eq:id}) yields the lower bound. These bounds were derived
by \citet{Rotnitzky2003}, \citet{Zhang2003}
and \citet{Hudgens2003}.
Consistent estimates of (\ref{eq:upper}) and (\ref{eq:lower}) can be
computed by replacing
$E[Y(0)|S(0)=1]$ with $\sum_i Y_i I(S_i=1,Z_i=0) / \sum_i
I(S_i=1,Z_i=0)$ and $\Prr[S(1)=1|S(0)=1]$ with
\[
\min \biggl\{ 1, \frac{
\sum_i I(S_i=Z_i=1)/\sum_i I(Z_i=1)
}{
\sum_i I(S_i=1,Z_i=0)/\sum_i I(Z_i=0)
} \biggr\}.
\]
Returning to the pertussis vaccine study, the estimated lower and upper
bounds of
(\ref{eq:ACEy1}) are $-$0.57 and $-$0.15. These estimated bounds exclude
zero, leading to the conclusion (ignoring
sampling variability) that vaccination lowers the risk of severe
pertussis in
individuals who will become infected regardless of whether they are vaccinated.

Note if $\Prr[S(1)=1|S(0)=1]=1$, that is, the vaccine has no protective effect
against infection, then the protected principal stratum
$S^{P_{0}}=(1,0)$ is
empty and both (\ref{eq:upper}) and (\ref{eq:lower}) equal $E[Y(0)|S(0)=1]$
meaning that (\ref{eq:ACEy1}) is identifiable and equals $E[Y|Z=1,S=1] -
E[Y|Z=0,S=1]$.
Intuitively, the lack of vaccine effect against infection eliminates
the potential for selection bias.

As discussed in Section~\ref{sec:covadjust}, incorporation of
covariates can tighten bounds. For
covariates $X$ with finite support, one simple approach of adjusting
for covariates entails
determining bounds within strata defined by the levels of $X$ and then
taking a
weighted average of the within strata bounds over the distribution of $X$.
For the bounds in (\ref{eq:upper}) and (\ref{eq:lower}), adjustment
for covariates will always
lead to bounds that are at least as tight as bounds unadjusted for
covariates (\cite{Lee2009}; \cite{Long2013}).

If the observed data provide evidence contrary to monotonicity (\ref
{eq:monos}), then bounds may be
obtained under only (\ref{eq:ignore}). Without monotonicity (\ref{eq:monos}),
the proportion of infected controls that are in the doomed principal
stratum is
no longer identified but may be bounded in order to arrive at bounds for
$E[Y(0)|S^{P_{0}}=(1,1)]$. In addition, the harmed principal stratum
defined by $S^{P_{0}} = (0, 1)$ is no longer empty and thus
$E[Y(1)|S^{P_{0}}=(1,1)]$
is no longer identifiable from the observed data and may also be
bounded in a similar fashion to $E[Y(0)|S^{P_{0}}=(1,1)]$. Details
regarding these bounds without the monotonicity assumption may be found in
\citet{Zhang2003} and \citet{Grilli2008}.

%s3.4 #&#
\subsection{Sensitivity Analysis}\label{sec:pesensanal}

The bounds (\ref{eq:upper}) and (\ref{eq:lower}) are useful in
bounding the
vaccine effect on $Y$ in the doomed stratum. However, these bounds
may be rather extreme. An alternative approach is
to make an untestable assumption that identifies the post-infection vaccine
effect on $Y$ and then consider how sensitive the resulting inference
is to
departures from this assumption. For instance, assuming
%
%e18 #&#
\begin{eqnarray}
\label{eq:noselect} &&\Prr\bigl[Y(0)=1 | S^{P_0}=(1,1)\bigr]
\nonumber
\\[-8pt]
\\[-8pt]
&&\quad = \Prr\bigl[Y(0)=1 | S^{P_0}=(1,0)\bigr],
\nonumber
\end{eqnarray}
identifies (\ref{eq:ACEy1}). \citet{Hudgens2006}
refer to this as the no selection model. To examine how inference varies
according to departures from (\ref{eq:noselect}), following
\citet{Scharfstein1999}, and \citet{Robinsetal1999}, consider the following
sensitivity parameter:
%
%e19 #&#
\begin{eqnarray}
\label{eq:selectionmod1} \exp(\gamma)&=&\bigl(\Prr\bigl[Y(0)=1 | S^{P_0}=(1,1)
\bigr]
\nonumber
\\
&&\hspace*{5pt} {}/\Prr\bigl[Y(0)=0 | S^{P_0}=(1,1)\bigr]\bigr)
\nonumber
\\[-8pt]
\\[-8pt]
&&{}\cdot\bigl(\Prr\bigl[Y(0)=1 | S^{P_0}=(1,0)\bigr]
\nonumber
\\
&&\hspace*{18pt} {}/\Prr\bigl[Y(0)=0 | S^{P_0}=(1,0)\bigr]
\bigr)^{-1}.
\nonumber
\end{eqnarray}
In words, $\exp(\gamma)$ compares the odds of severe disease when not
vaccinated
in the doomed versus the protected principal stratum. Assuming
(\ref{eq:noselect}) corresponds to $\gamma=0$. A sensitivity analysis entails
examining how inference about (\ref{eq:ACEy1}) varies as $\gamma$
becomes farther
from 0. For any fixed value of $\gamma$, (\ref{eq:ACEy1}) is
identified (see
Figure~\ref{fig:statsci1}) and can be consistently estimated by maximum likelihood
estimation without any
additional assumptions (\cite{Gilbert2003}).
The lower and upper bounds (\ref{eq:lower}) and (\ref{eq:upper}) are
obtained by letting $\gamma\to\infty$ and
$\gamma\to-\infty$. To see this, note that as $\gamma\to\infty$
(\ref{eq:selectionmod1}) implies in the limit that either
\begin{eqnarray*}
\Prr\bigl[Y(0)=1|S^{P_0}=(1,1)\bigr]&=&1\quad \mbox{or }
\\
\Prr\bigl[Y(0)=1|S^{P_0}=(1,0)\bigr]&=&0,
\end{eqnarray*}
which is equivalent to (\ref{eq:lowerc1}). Sensitivity analysis can be
conducted
by letting $\gamma$ range over a set of values~$\Gamma$.

Tighter bounds can be achieved by placing restrictions on $\Gamma$,
perhaps based on prior beliefs
about $\gamma$ elicited from subject matter experts. For example,
\citet{shepherd2007}
surveyed 10 recognized HIV experts in order to elicit a plausible range
for a
sensitivity parameter representing a departure from the assumption of
no selection bias between
vaccinated and unvaccinated individuals who acquired HIV during an HIV
vaccine trial. Included in this survey
was the analysis approach, a brief explanation of the potential
for selection bias, the definition of the sensitivity parameter being
employed, examples of the implications of
certain sensitivity parameter values on selection bias and possible
justification for believing certain values of the
sensitivity parameter. The expert responses to the survey were fairly
consistent and several written
justifications for the respondents' chosen ranges indicated a high
level of understanding of both the counterfactual
nature of the sensitivity parameter and the need to account for
selection bias.
\nocite{shepherd2007}

%%%%%%%%%%%%%%%%%%%%%%%%%%%%%%%%%%%%%%%%%%%%%%%%%%%%%%%%%%%%%%%%%%%%%%%%%%%%%%%%%%%%%%%%%%%%%%%%%%%%%%%%%%%%%%%%%%%
%%%%%%%%%%%%%%%%%%%%%%%%%%%%%%%%%%%%%%%%%%%%%%%%%%%%%%%%%%%%%%%%%%%%%%%%%%%%%%%%%%%%%%%%%%%%%%%%%%%%%%%%%%%%%%%%%%%
%%%%%%%%%%%%%%%%%%%%%%%%%%%%%%%%%%%%%%%%%%%%%%%%%%%%%%%%%%%%%%%%%%%%%%%%%%%%%%%%%%%%%%%%%%%%%%%%%%%%%%%%%%%%%%%%%%%
%%%%%%%%%%%%%%%%%%%%%%%%%%%%%%%%%%%%%%%%%%%%%%%%%%%%%%%%%%%%%%%%%%%%%%%%%%%%%%%%%%%%%%%%%%%%%%%%%%%%%%%%%%%%%%%%%%%
%%%%%%%%%%%%%%%%%%%%%%%%%%%%%%%%%%%%%%%%%%%%%%%%%%%%%%%%%%%%%%%%%%%%%%%%%%%%%%%%%%%%%%%%%%%%%%%%%%%%%%%%%%%%%%%%%%%
%%%%%%%%%%%%%%%%%%%%%%%%%%%%%%%%%%%%%%%%%%%%%%%%%%%%%%%%%%%%%%%%%%%%%%%%%%%%%%%%%%%%%%%%%%%%%%%%%%%%%%%%%%%%%%%%%%%

%s4 #&#
\section{Randomized Studies with Partial Compliance}\label{sec:compliance}
%s4.1 #&#
\subsection{Global Average Treatment Effect}\label{sec:GATE}

In a placebo controlled randomized trial where (\ref{eq:ign}) holds
but there is
non-compliance (i.e., individuals are randomly assigned to treatment or control
but they do not necessarily adhere or comply with their assigned treatment),
the naive estimator is a consistent estimator of the average effect of
treatment \itshape assignment. \normalfont However, in this case
parameters other than
the effect of treatment assignment may be of interest. As in the last
section, a principal
effect may be defined using compliance as the intermediate post-randomization
variable over which to define principal strata; namely the principal strata
would consist of individuals who would comply with their randomization
assignment if assigned treatment or control or ``compliers,''
individuals who
would always take treatment regardless of randomization or ``always takers,''
individuals who never take treatment ``never takers'' and individuals
who take
treatment only if assigned control or ``defiers.'' A principal effect
of interest might be the effect of treatment in the
complier principal stratum (\cite{imbens1994}; \cite{Angrist1996}), in which
case bounds and sensitivity
analyses similar to those in Section~\ref{sec:princstrat} are
applicable. However, as several authors
including \citet{Robins1989} and \citet{Robins1996} have pointed out,
such principal effects may not be of ultimate public health interest because
they only apply to the subpopulation of compliers in clinical trials, which
may differ from the population that elect to take treatment once licensed.
For example, once efficacy is proved, a larger subpopulation of people
may be
willing to take the treatment. Effects defined on the subpopulation of
compliers are
also of limited decision-making utility because individual principal
stratum membership
is generally unknown prior to treatment assignment (\cite{Joffe2011}).

\citet{Robins1996} suggested that in settings where
the trial population could be persuaded to take the treatment once
licensed, a
more relevant public health estimand is the global average treatment effect,
defined as the average effect of actually taking treatment versus not
taking treatment given treatment assignment $z$. This causal estimand
is similar
to the average treatment effect defined in Section~\ref{sec:unmeasconfound}, but requires generalizing
the potential outcome definitions used previously to include separate potential
outcomes for each of the four combinations of treatment assignment and
actual treatment received.
For further discussion regarding causal models in presence of
noncompliance, see
\citet{Chickering1996} and \citet{Dawid2002} among others.

Suppose we observe data from a clinical trial where each individual is
randomly assigned to treatment or control.
Let $Z$ indicate treatment assignment where $Z = 1$ denotes treatment
and $Z = 0$ denotes control. Suppose individuals do
not necessarily comply with their randomization assignment and let $S$
be a variable indicating whether or not treatment
was actually taken, where $S = 1$ denotes treatment was taken and $S =
0$ otherwise. Thus, an individual is compliant with
their randomization assignment if $S=Z$. Let $Y$ be a binary outcome of
interest. Denote the potential treatment taken by
$S(z)$ for $z = 0, 1$, where $S(z) = 1$ indicates taking treatment when
assigned  $z$ and $S(z) = 0$ denotes not taking
treatment when assigned $z$. Let $Y (z, s)$ denote the potential
outcome if an individual is assigned treatment $z$ but actually
takes treatment $s$. Conceiving of these potential outcomes
depends on a supposition that trial participants who did not comply in
the trial
could be persuaded to take the treatment under other circumstances.
Given this supposition, the global average treatment effect for each treatment
assignment $z=1$ and $z=0$ is defined as $\GATE_z = E[Y(z,1) - Y(z,0)]$.
For instance, $\GATE_1$ is the difference in the average outcomes under the
counterfactual scenario everyone was assigned vaccine and did comply versus
the counterfactual scenario everyone was assigned vaccine but did not comply.

Bounds for $\GATE_z$ are given below under three assumptions:
independent treatment
assignment
%
%e20 #&#
\begin{eqnarray}
\label{eq:ignore1} &&Z \amalg\bigl\{S(0),S(1),Y(0,0),
\nonumber
\\[-8pt]
\\[-8pt]
&&\hspace*{25pt}Y(0,1),Y(1,0),Y(1,1)\bigr\};
\nonumber
\end{eqnarray}
monotonicity with respect to $S$
%
%e21 #&#
\begin{equation}
\label{eq:monos1} \Prr\bigl[S(1) \geq S(0)\bigr] = 1;
\end{equation}
and the exclusion restriction
%
%e22 #&#
\begin{equation}
\label{eq:er} Y(0,s)=Y(1,s) \quad \mbox{for } s=0,1.
\end{equation}
Assumption (\ref{eq:er})
indicates treatment assignment has no effect when the actual treatment
taken is
held fixed. Under (\ref{eq:er}), $\GATE_0 =\GATE_1$ which we denote
by GATE. In
this case each individual has two potential outcomes according to $s=0$ and
$s=1$ [which could be denoted by $Y(s) = Y(0,s)= Y(1,s)$ for $s=0,1$]
and GATE is
equivalent to the ATE discussed in Section~\ref{sec:unmeasconfound}
with $z$ replaced by $s$. \citet{Robins1989}
derived bounds for GATE under several different combinations of
(\ref{eq:ignore1})--(\ref{eq:er})
as well as some
additional assumptions such as monotonicity with respect to $S$, that
is, $Y(z,1)
\geq Y(z,0)$ for $z=0,1$.
\citet{Manski1990} independently derived related results.
Under (\ref{eq:ignore1})--(\ref{eq:er}),
the sharp lower and upper bounds on GATE are
%
%e23 #&#
\begin{eqnarray}
\label{eq:lgate} &&-1 +\max_z \bigl\{\Prr[Y=1,S=1|Z=z]\bigr\}
\nonumber
\\[-8pt]
\\[-8pt]
&&\quad{} + \max_z \bigl\{\Prr[Y=0,S=0|Z=z]\bigr\}
\nonumber
\end{eqnarray}
and
%
%e24 #&#
\begin{eqnarray}
\label{eq:ugate} &&1 -\max_z\bigl\{\Prr[Y=0,S=1|Z=z]\bigr\}
\nonumber
\\[-8pt]
\\[-8pt]
&&\quad{} -\max_z \bigl\{\Prr[Y=1,S=0|Z=z]\bigr\}.
\nonumber
\end{eqnarray}

\citet{Balke1997} derived sharp
bounds for GATE under a variety of assumptions, including (\ref
{eq:ignore1})--(\ref{eq:er}),
by recognizing that the derivation of the bounds is equivalent to a
linear programming optimization problem.
To see that bounds can be formulated as a linear programming optimization
problem, first note that GATE can be expressed as a linear combination of
probabilities of the joint distribution of $\boldL = (Y(0,0),
Y(0,1), Y(1,0), Y(1,1), S(0),\break
S(1))$
%
%e25 #&#
\begin{equation}
\label{eq:objftn} \sum_{ l_1 \in\mathcal{L}_1} \Prr[\boldL=l_1]
- \sum_{ l_0 \in\mathcal{L}_0} \Prr[\boldL=l_0],
\end{equation}
where $\mathcal{L}_s$ is the set of possible realizations of $L$ where
$Y(0,s)=Y(1,s)=1$ for $s=0,1$.
Under independent treatment assignment, there exists a linear
transformation between the probabilities in the joint
distribution of $L$ and the probabilities in the conditional
distribution of the observable
random variables $Y$ and $S$ given $Z$, namely
%
%e26 #&#
\begin{equation}
\label{eq:obsconstraint} \Prr[Y=y,S=s|Z=z] = \sum_{ l\in\mathcal{O}_{ys\cdot z}} \Prr[
\boldL=l],
\end{equation}
where $\mathcal{O}_{ys\cdot z}$ is the set of possible realizations of
$\boldL$
where $S(z)=s$ and $Y(z,s)=y$ for $z,y,s=0,1$.
To find the sharp bounds, the objective function (\ref{eq:objftn}) is
minimized (or maximized) subject to the constraints
(\ref{eq:obsconstraint}), $\Prr[L=l]\geq0$ for every $l\in\mathcal
{L}$, and $\sum_{l\in\mathcal{L}}
\Prr[\boldL=l] = 1$ where $\mathcal{L}$ is the set of all
possible realizations of $\boldL$
assuming (\ref{eq:monos1}) and (\ref{eq:er}). Optimization may
be accomplished using the simplex algorithm and the dimension of this
problem permits obtaining a closed form solution involving
probabilities of the observed data distribution (\cite{Balke1993}),
namely (\ref{eq:lgate}) and (\ref{eq:ugate}).
\nocite{Balke1993}

If in addition to assumptions (\ref{eq:ignore1}) and (\ref{eq:er}),
it is assumed that
%
%e27 #&#
\begin{eqnarray}
\label{eq:noeffectmod} &&E\bigl[Y(z,1)-Y(z,0)|Z=1,S=s\bigr]
\nonumber
\\[-8pt]
\\[-8pt]
&&\quad=E\bigl[Y(z,1)-Y(z,0)|Z=0,S=s\bigr]
\nonumber
\end{eqnarray}
for $s,z=0,1$ then GATE is identified and equals
%
%e28 #&#
\begin{equation}
\label{eq:ivestimand} \frac{E[Y|Z=1]-E[Y|Z=0]}{E[S|Z=1]-E[S|Z=0]}
\end{equation}
(\cite{Hernan2006}). For $s=0$ assumption (\ref{eq:noeffectmod}) is
known as a no current treatment interaction
assumption (\cite{Robins1994}), and expression (\ref{eq:ivestimand})
is known as the instrumental variables estimand
(\cite{imbens1994}; \cite{Angrist1996}). Sensitivity analyses may be conducted
by defining sensitivity
parameters representing departures from (\ref{eq:ignore1}), (\ref
{eq:er}) or (\ref{eq:noeffectmod}) and
then examining how inference about GATE varies as values of these
parameters change. For instance, \citet{Robinsetal1999}
define current treatment interaction functions which represent a
departure from (\ref{eq:noeffectmod}) for $s=0$.

%s4.2 #&#
\subsection{Cholestyramine Example}\label{sec:cholestyramine}

To illustrate the GATE, we consider data presented in
\citeauthor{Pearl2009} (\citeyear{Pearl2009}, Section~8.2.6)
on 337 subjects who participated in a randomized trial to assess the
effect of
cholestyramine on cholesterol reduction. Let $Z=1$ denote
assignment to cholestyramine and $Z=0$ assignment to placebo. Let $S=1$ if
cholestyramine was actually taken by the participant and $S=0$
otherwise. Let
$Y=1$ if the participant had a response and $Y=0$ otherwise, where
response is
defined as reduction in the level of cholesterol by 28 units or more. Pearl
reported the following observed proportions:
\begin{eqnarray*}
\hatPr[Y=0,S=0|Z=0]&=&0.919,
\\
\hatPr[Y=0,S=0|Z=1]&=&0.315,
\\
\hatPr[Y=0,S=1|Z=0]&=&0.000,
\\
\hatPr[Y=0,S=1|Z=1]&=&0.139,
\\
\hatPr[Y=1,S=0|Z=0]&=&0.081,
\\
\hatPr[Y=1,S=0|Z=1]&=&0.073,
\\
\hatPr[Y=1,S=1|Z=0]&=&0.000,
\\
\hatPr[Y=1,S=1|Z=1]&=&0.473.
\end{eqnarray*}
No participants assigned placebo actually took choles\-tyramine,
suggesting
the monotonicity assumption (\ref{eq:monos1}) is reasonable.
On the other hand, 38.8\% of individuals assigned treatment did not
actually take
cholestyramine.

From (\ref{eq:lgate}) and (\ref{eq:ugate}), the bounds on GATE assuming
(\ref{eq:monos1}), (\ref{eq:ignore1}) and
(\ref{eq:er})
are estimated to be $-1+\max\{0.000,0.473\} + \max\{0.919,0.315\} =
0.392$ and
$1 - \max\{0, 0.139\} - \max\{0.081,0.073\} = 0.780$. The positive
sign of the
estimated bounds indicates the treatment is beneficial. Pearl
interprets the
estimated bounds as follows: ``although 38.8\% of the subjects deviated from
their treatment protocol, the experimenter can categorically state
that, when
applied uniformly to the population, the treatment is guaranteed to
increase by
at least 39.2\% the probability of reducing the level of cholesterol by
28 points
or more.'' Such an interpretation does not account for sampling
variability, the
topic of Section~\ref{sec:irur}.

%%%%%%%%%%%%%%%%%%%%%%%%%%%%%%%%%%%%%%%%%%%%%%%%%%%%%%%%%%%%%%%%%%%%%%%%%%%%%%%%%%%%%%%%%%%%%%%%%%%%%%%%%%%%%%%%%%%
%%%%%%%%%%%%%%%%%%%%%%%%%%%%%%%%%%%%%%%%%%%%%%%%%%%%%%%%%%%%%%%%%%%%%%%%%%%%%%%%%%%%%%%%%%%%%%%%%%%%%%%%%%%%%%%%%%%
%%%%%%%%%%%%%%%%%%%%%%%%%%%%%%%%%%%%%%%%%%%%%%%%%%%%%%%%%%%%%%%%%%%%%%%%%%%%%%%%%%%%%%%%%%%%%%%%%%%%%%%%%%%%%%%%%%%
%%%%%%%%%%%%%%%%%%%%%%%%%%%%%%%%%%%%%%%%%%%%%%%%%%%%%%%%%%%%%%%%%%%%%%%%%%%%%%%%%%%%%%%%%%%%%%%%%%%%%%%%%%%%%%%%%%%
%%%%%%%%%%%%%%%%%%%%%%%%%%%%%%%%%%%%%%%%%%%%%%%%%%%%%%%%%%%%%%%%%%%%%%%%%%%%%%%%%%%%%%%%%%%%%%%%%%%%%%%%%%%%%%%%%%%
%%%%%%%%%%%%%%%%%%%%%%%%%%%%%%%%%%%%%%%%%%%%%%%%%%%%%%%%%%%%%%%%%%%%%%%%%%%%%%%%%%%%%%%%%%%%%%%%%%%%%%%%%%%%%%%%%%%

%s5 #&#
\section{Mediation Analysis}\label{sec:mediation}
%s5.1 #&#
\subsection{Natural Direct and Indirect Effects}\label{sec:natural}
As demonstrated in Sections~\ref{sec:princstrat} and \ref
{sec:compliance}, independent treatment assignment does not
guarantee that the causal estimand of interest will be identifiable.
Another setting where this occurs is in mediation analysis,
where researchers are interested in
whether or not the effect of a treatment is mediated by some intermediate
variable. Even in studies where treatment is assigned randomly and
there is
perfect compliance, confounding may exist between the intermediate
variable and
the outcome of interest such that effects describing the mediated relationships
will not in general be identifiable. Thus, bounds and sensitivity analysis
may be helpful in drawing inference.

To illustrate, let $Y$ be an observed binary outcome of interest, and
$S$ a
binary intermediate variable observed some time between treatment assignment
$Z$ and the observation of $Y$. The goal is to assess whether and to
what extent
the effect of $Z$ on $Y$ is mediated by or through $S$. Denote the potential
outcome of the intermediate variable under treatment $z$ by $S(z)$ for $z=0,1$
such that $S=S(Z)$, and the potential outcomes under treatment $z$ and
intermediate $s$ as $Y(z, s)$ such that $Y=Y(Z,S(Z))$. Here, as in the previous
section, it is assumed that both $Z$ and $S$ can be set to particular fixed
values, such that there are four potential outcomes for $Y$ per individual.
Unless otherwise specified, independent treatment assignment (\ref
{eq:ignore1}) will be
assumed throughout this section.

Define the total effect of treatment to be $E[Y(1,\allowbreak  S(1))-Y(0,S(0))]$,
which is
equivalent to the ATE defined in Section~\ref{sec:noassbounds}.
\nocite{Pearl2001} The total effect
of treatment can be decomposed in the following way:
%
%e29 #&#
\begin{eqnarray}
\label{eq:decompdirectindirect} %
&&E\bigl[Y\bigl(1, S(1)\bigr)-Y\bigl(0,S(0)\bigr)\bigr]
\nonumber
\\
&&\quad = E\bigl[Y\bigl(1,S(z)\bigr)-Y\bigl(0,S(z)\bigr)\bigr]
\\
&&\qquad {} + E\bigl[Y\bigl(z', S(1)\bigr)-Y(z',S(0)
\bigr]
\nonumber
\end{eqnarray}
for $z=0,1$ and $z'=1-z$. The right-hand side of (\ref
{eq:decompdirectindirect}) decomposes the
total effect into the sum of two separate effects. The first
expectation on the
right-hand side of (\ref{eq:decompdirectindirect}) is the natural
direct effect for
treatment $z$, $\NDE_z= E[Y(1,S(z))-Y(0,S(z))]$
(\cite{Robins1992}; \cite{Pearl2001}; \cite{Robins2003}; \cite{Kaufman2009}; \cite{Robins2010}).
The natural direct
effect is the average effect of the treatment on the outcome when the
intermediate variable is set to the potential value that would occur under
treatment assignment $z$. The second expectation on the right-hand side of
(\ref{eq:decompdirectindirect}) is the natural indirect effect,
$\NIE_{z}=E[Y(z,S(1))-Y(z,S(0))]$ (\cite{Pearl2001}; \cite{Robins2003}; \cite{Imai2010a}).
The natural indirect effect is
the difference in the average outcomes when treatment is set to $z$ and the
intermediate variable is set to the value that would have occurred under
treatment compared to if the intermediate variable were set to the
value that
would have occurred under control.

Though the total effect is identifiable assuming (\ref{eq:ignore1}),
the natural
direct and indirect effects are not identifiable since they entail
$E[Y(z,S(1-z))]$ which depends on unobserved counterfactual distributions.
\citet{Sjolander2009a} derived bounds for the natural
direct effects assuming only independent treatment assignment (\ref
{eq:ignore1})
using the linear programming technique of \citet{Balke1997}. This
results in the following sharp lower and upper bounds for $\NDE_0$ and $\NDE_1$:
%
%e30 #&#
%e31 #&#
\begin{eqnarray}
\label{eq:boundsNDE0} &&\max\lleft\{ %
\begin{array} {c} -p_{11\cdot0}-p_{10\cdot0},
\\
p_{11\cdot1}+p_{01\cdot0}-1-p_{10\cdot0},
\\
p_{10\cdot1} +p_{00\cdot0} -1 -p_{11\cdot0} \end{array} %
 \rright\}
\nonumber
\\[-8pt]
\\[-8pt]
&&\quad \leq\NDE_0 \leq\min\lleft\{ %
\begin{array}
{c}p_{01\cdot0}+p_{00\cdot0},
\\
1-p_{00\cdot1} +p_{01\cdot0}-p_{10\cdot0},
\\
1-p_{01\cdot1} +p_{00\cdot0}-p_{11\cdot0} \end{array} %
 \rright\},
\nonumber
\\
\label{eq:boundsNDE1} &&\max\lleft\{ %
\begin{array} {c}-p_{01\cdot1}-p_{00\cdot1},
\\
p_{00\cdot0}-1 -p_{01\cdot1}+p_{10\cdot1},
\\
p_{01\cdot0}-1 -p_{00\cdot1}+p_{11\cdot1} \end{array} %
 \rright\}
\nonumber
\\[-8pt]
\\[-8pt]
&&\quad \leq\NDE_1 \leq\min\lleft\{ %
\begin{array} {c}
p_{11\cdot1}+p_{10\cdot1},
\\
1-p_{01 \cdot1}+p_{10\cdot1} -p_{11\cdot0},
\\
1-p_{00\cdot1}+p_{11\cdot1}-p_{10\cdot0} \end{array} %
 \rright\},
\nonumber
\end{eqnarray}
where $p_{ys\cdot z}=\Prr(Y=y,S=s|Z=z)$.
These bounds may exclude 0, indicating a natural direct effect of
treatment $z$ when the intermediate
variable is set to $S(z)$ (ignoring sampling variability). There are
instances where the bounds in
(\ref{eq:boundsNDE0}) and (\ref{eq:boundsNDE1}) may collapse to a
single point, for example, if
$p_{10\cdot0}=p_{10\cdot1}=1$. Using (\ref{eq:decompdirectindirect}),
bounds for $\NIE_0$ and $\NIE_1$ can be obtained by subtracting the
bounds for $\NDE_1$ and $\NDE_0$ from the
total effect, which is identified under (\ref{eq:ignore1}) and equal to
$(p_{11\cdot1}+p_{10\cdot1})-(p_{10\cdot0} - p_{11\cdot0})$.

Just as in Sections~\ref{sec:unmeasconfound}--\ref{sec:compliance},
monotonicity assumptions can be made to tighten the
above bounds. For instance, if
\begin{eqnarray*}
\Prr\bigl[S(0)\leq S(1) \bigr]&=&1,
\\
\Prr\bigl[Y(0,s)\leq Y(1,s)\bigr]&=&1\quad \mbox{for }s=0,1\quad \mbox{and}
\\
\Prr\bigl[Y(z,0)\leq Y(z,1)\bigr]&=&1\quad \mbox{for }z=0,1,
\end{eqnarray*}
are assumed, then $\Prr[\boldL=l]=0$ for all $l$ such that (i)~$S(0)=1$ and $S(1)=0$, (ii) $Y(0,s)=1$ and $Y(1,s)=0$ for $s=0$ or $1$
or (iii) $Y(z,0)=1$ and $Y(z,1)=0$ for $s=0$ or $1$, which restricts
the feasible region of the
linear programming problem. The resulting sharp bounds for the natural
direct effect are
%
%e32 #&#
\begin{eqnarray}
\label{eq:boundsNDEmono} &&\max\lleft\{ %
\begin{array} {c} 0,
p_{01\cdot0}-p_{01\cdot1}, p_{10\cdot
1}-p_{10\cdot0},
\\
p_{01\cdot0}-p_{01\cdot1}+p_{10\cdot1}-p_{10\cdot0} \end{array}
 \rright\}
\nonumber
\\[-8pt]
\\[-8pt]
&&\quad \leq\NDE_z \leq p_{10\cdot1}+p_{11\cdot1}
-p_{10\cdot0}-p_{11\cdot0}
\nonumber
\end{eqnarray}
(\cite{Sjolander2009a}). The bounds (\ref{eq:boundsNDEmono}) are
always at least as
narrow as (\ref{eq:boundsNDE0}) and (\ref{eq:boundsNDE1}).
Interestingly these
narrower bounds do not depend on $z$. The bounds in
(\ref{eq:boundsNDEmono}) may also collapse to a single point, for example, if
$p_{10\cdot0}=p_{10\cdot1}$ and $p_{01\cdot0}-p_{01\cdot1}
 = p_{11\cdot1} -p_{11\cdot0}$.

The natural direct effect provides insight into whet\-her or not
treatment yields
additional benefit on the outcome of interest when the influence of
treatment on
the intermediate variable is eliminated. However, researchers might
also be
interested in what benefit is provided by treatment if the effect of
the intermediate variable on the outcome is eliminated or held constant.
This question suggests a different causal estimand known as the
controlled direct
effect. Bounds for the controlled direct effect can be found in
\citet{Pearl2001}, \citet{Cai2008}, \citet{Sjolander2009a} and \citet{VanderWeele2011e}.

%s5.2 #&#
\subsection{Sensitivity Analysis}\label{sec:medsensanal}
As in other settings where the effect of interest is not identifiable,
sensitivity analysis in the mediation setting may be conducted by making
untestable assumptions that identify the direct or indirect effects. Then
sensitivity of inference to departures from these assumptions can be examined.
For example, if (\ref{eq:ignore1}) holds, then the natural direct and indirect
effects are identified under the following additional assumptions
%
%e33 #&#
%e34 #&#
\begin{eqnarray}
\label{eq:identnatural1} &&Y(z,s) \amalg S|Z\quad \mbox{for }z,s=0,1\quad \mbox{and }
\\
\label{eq:identnatural2} &&Y(z,s) \amalg S\bigl(z'\bigr) \quad \mbox{for }
z,z',s=0,1
\end{eqnarray}
(\cite{Pearl2001}; \cite{VanderWeele2010}).
Assumption (\ref{eq:identnatural1})
would be valid if subjects were randomly assigned $S$ within different
levels of treatment assignment $Z$.
In settings where $S$ is not randomly assigned,
(\ref{eq:identnatural1}) might be considered plausible if it is
believed that conditional on $Z$
there are no variables which confound the mediator--outcome relationship.
Both assumptions (\ref{eq:identnatural1}) and (\ref{eq:identnatural2})
will not hold in general if $Z$ has an effect on some other
intermediate variable, say $R$, which in turn has an
effect on both $S$ and $Y$. Thus, (\ref{eq:identnatural1}) and (\ref
{eq:identnatural2}) may fail unless
the mediator $S$ occurs shortly after treatment~$Z$.
Under assumptions (\ref{eq:ignore1}), (\ref{eq:identnatural1}) and
(\ref{eq:identnatural2}),
\begin{eqnarray*}
\NDE_z &=& (-1)^z\sum_s
\bigl\{E[Y|Z=1-z,S=s]
\\
&&{}-E[Y|Z=z,S=s]\bigr\}\Prr[S=s|Z=z]
\end{eqnarray*}
and
\begin{eqnarray*}
&&\NIE_z = (-1)^z\sum_s
E[Y|Z=z,S=s]
\\
&&\hspace*{78pt} {}\cdot\bigl\{\Prr[S=s|Z=1-z]
\\
&&\hspace*{93pt} {}-\Prr[S=s|Z=z]\bigr\}.
\end{eqnarray*}
Because assumptions (\ref{eq:identnatural1})
and (\ref{eq:identnatural2}) cannot be empirically tested,
sensitivity analysis should be
conducted. Similar to Section~\ref{sec:unmeasconfoundsensanal},
sensitivity analysis might proceed by positing the
existence of an unmeasured confounding variable $U$ associated with the
potential
mediator values $S(z)$ and the potential outcomes $Y(z,s)$ for
$z,s=0,1$. Assumption
(\ref{eq:identnatural1}) would then replaced by $Y(z,s) \amalg S|\{Z,U\}$ and
(\ref{eq:identnatural2}) by $Y(z,s) \amalg S(z')|U$ for
$s,z,z'=0,1$. Sensitivity analysis would then
proceed by exploring how inference about the natural direct and
indirect effects changes as the magnitude of the
associations of $U$ with $S(z)$ and $Y(s,z')$ for $z,z',s = 0,1$
vary. For further
details regarding bounds and sensitivity analysis in mediation
analysis, see
\citet{Imai2010a}, \citet{VanderWeele2010} and \citet{Hafeman2011}.

%%%%%%%%%%%%%%%%%%%%%%%%%%%%%%%%%%%%%%%%%%%%%%%%%%%%%%%%%%%%%%%%%%%%%%%%%%%%%%%%%%%%%%%%%%%%%%%%%%%%%%%%%%%%%%%%%%%
%%%%%%%%%%%%%%%%%%%%%%%%%%%%%%%%%%%%%%%%%%%%%%%%%%%%%%%%%%%%%%%%%%%%%%%%%%%%%%%%%%%%%%%%%%%%%%%%%%%%%%%%%%%%%%%%%%%
%%%%%%%%%%%%%%%%%%%%%%%%%%%%%%%%%%%%%%%%%%%%%%%%%%%%%%%%%%%%%%%%%%%%%%%%%%%%%%%%%%%%%%%%%%%%%%%%%%%%%%%%%%%%%%%%%%%
%%%%%%%%%%%%%%%%%%%%%%%%%%%%%%%%%%%%%%%%%%%%%%%%%%%%%%%%%%%%%%%%%%%%%%%%%%%%%%%%%%%%%%%%%%%%%%%%%%%%%%%%%%%%%%%%%%%
%%%%%%%%%%%%%%%%%%%%%%%%%%%%%%%%%%%%%%%%%%%%%%%%%%%%%%%%%%%%%%%%%%%%%%%%%%%%%%%%%%%%%%%%%%%%%%%%%%%%%%%%%%%%%%%%%%%
%%%%%%%%%%%%%%%%%%%%%%%%%%%%%%%%%%%%%%%%%%%%%%%%%%%%%%%%%%%%%%%%%%%%%%%%%%%%%%%%%%%%%%%%%%%%%%%%%%%%%%%%%%%%%%%%%%%

%s6 #&#
\section{Longitudinal Treatment}\label{sec:longtrt}
%s6.1 #&#
\subsection{Background} \label{sec:backgroundlong}
In Sections~\ref{sec:unmeasconfound}--\ref{sec:mediation}, treatment
is assumed to remain fixed across follow up time and
outcomes are one-dimensional. However, frequently researchers are
interested in
assessing causal effects comparing longitudinal outcomes for patients on
different treatment regimens where treatment may vary in time. As the number
of times at which an individual may receive treatment increases, the
number of
possible treatment regimens increases exponentially. Because each treatment
regimen corresponds to a separate potential (longitudinal) outcome and
only one
potential outcome is ever observed, the fraction of potential outcomes
that are
unobserved quickly grows close to one as the number of possible
treatment times
increases. As in other settings, unless treatment regimens are randomly
assigned, regimen effects will not be identifiable without additional
assumptions. In the longitudinal setting, bounds will typically be largely
uninformative because of the high proportion of unobserved potential outcomes.
Therefore, analyses usually proceed by invoking modeling assumptions
that render
treatment effects identifiable and then conducting sensitivity analysis
corresponding to key untestable modeling assumptions.
\nocite{Vanderlaan2003}

Models for potential outcomes as functions of covariates (such as
treatment) and
possibly other potential outcomes are often referred to as structural models.
For longitudinal potential outcomes and treatments, popular models
include structural nested models and marginal structural models
(\cite{Robinsetal1999}; \cite{Robins1999}; \cite{Vanderlaan2003}; \cite{Brumback2004}).
In Section~\ref{sec:msm}
below, we consider a marginal structural model where the treatment effect
is identified assuming conditionally independent treatment assignment.
Sensitivity analyses exploring departures from this assumption are then
considered in Section~\ref{sec:msmsensanal}.

%s6.2 #&#
\subsection{Marginal Structural Model}\label{sec:msm}
Consider a study where individuals possibly receive treatment at $\tau
$ fixed
time points (i.e., study visits). In general let
$\bar{A}(t)=(A(0),\ldots,A(t))$ represent the history of variable $A$
up to time
$t$ and $\bar{A}$ be the entire history of variable $A$ such that
$\bar{A} =
\bar{A}(\tau)$. Let $z(t)=1$ indicate treatment at visit $t$, and
$z(t)=0$ otherwise such that
$\bar{z}$ represents a treatment regimen for visits $0,\ldots, \tau$.
Denote the observed treatment regimen up to time $t$ as $\bar{Z}(t)$.
Let $Y$ be
some outcome of interest that may be categorical or continuous, and
denote the
potential outcome of $Y$ at visit $t$ for regimen $\bar{z}$ by $Y(\bar
{z},t)$ and the observed
outcome by $Y(t)$. Let
$\bar{X}(t)$ denote the history of some set of time varying covariates
up to time $t$, where
$X(0)$ denotes the baseline covariates.
Assume for simplicity there is no loss to follow-up or noncompliance
such that
we observe $n$ i.i.d. copies of $(\bar{Z}, \bar{Y}, \bar{X})$.

Consider the following marginal structural model of the mean potential outcome
were the entire population to follow regimen $\bar z$ up to time $t$:
%
%e35 #&#
\begin{eqnarray}
\label{eq:msm1} &&g\bigl(E\bigl[Y(\bar{z},t)|X(0)=x(0)\bigr]\bigr)
\nonumber
\\[-8pt]
\\[-8pt]
&&\quad =\beta_0+\beta_1\cum\bigl[\bar {z}(t-1)\bigr]+
\beta_2t + \beta_3 x(0)
\nonumber
\end{eqnarray}
for $t\in\{1,\ldots,\tau\}$, where $\cum[\bar{z}(t-1)]=\sum_{k=1}^{t-1}z(k)$ and $g(\cdot)$ is an
appropriate link function.
The causal estimand of interest is $\beta_1$, the regression
coefficient for
$\cum[\bar{z}(t-1)]$, which is the effect of having received treatment
at one
additional visit prior to time $t$ conditional on baseline covariates $X(0)$.
Because (\ref{eq:msm1}) involves counterfactual
outcome distributions, $\beta_1$ is not identifiable without additional
assumptions. One additional assumption is conditionally independent treatment
assignment
%
%e36 #&#
\begin{eqnarray}
\label{eq:seqignore}
Y(\bar{z},t)\amalg Z(k)|\bigl\{\bar{Z}(k-1),\bar{X}(k)\bigr\}
\nonumber
\\[-8pt]
\\[-8pt]
\eqntext{\mbox{for all } \bar{z} \mbox{ and } t > k}
\end{eqnarray}
(\cite{Robinsetal1999}; \cite{Robins1999}; \cite{Brumback2004}).
This assumption is true if the potential outcome at visit $t$ under treatment
regimen $\bar{z}$ is independent of the observed treatment at visit
$k$ given the history
of treatment up to visit $k-1$ and the covariate history up to visit $k$.
Assuming both a correctly specified model (\ref{eq:msm1}) and conditionally
independent treatment assignment (\ref{eq:seqignore}), fitting the
following model
to the observed data:
\begin{eqnarray*}
&&g\bigl(E\bigl[Y(t)|\bar{Z}(t-1)=\bar{z}(t-1),X(0)=x(0)\bigr]\bigr)
\\
&&\quad =\eta_0+\eta _1\cum\bigl[\bar{z}(t-1)\bigr]+
\eta_2t + \eta_3 x(0) ,
\end{eqnarray*}
using generalized estimating equations with an independent working
correlation matrix and time varying inverse probability of treatment
weights (IPTW) yields an estimator $\hat{\eta}_1$ that is consistent
for $\beta_1$
(\citeauthor{TchetgenTchetgen2012a}, \citeyear{TchetgenTchetgen2012a}, \citeyear{TchetgenTchetgen2012}).

%s6.3 #&#
\subsection{Sensitivity Analysis}\label{sec:msmsensanal}
If assumption (\ref{eq:seqignore}) does not hold, then the IPTW
estimator $\hat\eta_1$ is not necessarily consistent. Because
(\ref{eq:seqignore}) is not testable from the observed data,
sensitivity analysis might be considered to assess robustness of
inference to departures from (\ref{eq:seqignore}). Following \citet{Robins1999} and \citet{Brumback2004}, let
\begin{eqnarray*}
&& c\bigl(t,k,\bar{z}(t-1), \bar{x}(k)\bigr)
\\
&&\quad =E\bigl[Y(\bar{z},t)|\bar{Z}(k)=\bar {z}(k),\bar{X}(k)=\bar{x}(k) \bigr]
\\
&&\qquad {}- E\bigl[Y(\bar{z},t)|Z(k)=1-z(k),
\\
&&\hspace*{47pt}\bar{Z}(k-1)=\bar{z}(k-1),\bar {X}(k)=\bar{x}(k)\bigr]
\end{eqnarray*}
for $t > k$ and $\bar{z}$ such that $\Prr[Z(k)=z(k)|\bar{Z}(k-1)=\bar
{z}(k-1)]$ is
bounded away from 0 and 1. The function $c$ quantifies departures from
the conditional independent
treatment assignment assumption (\ref{eq:seqignore}) at each visit $t
> k$, where $c(t,k,\bar{z}(t-1),\bar{x}(k))=0$ for all
$\bar{z}$ and $t > k$ if (\ref{eq:seqignore}) holds. For the identity link, a bias
adjusted estimator of the causal effect $\beta_1$
may be obtained by recalculating the IPTW estimator with the observed
outcome $Y(t)$ replaced by $Y^\gamma(t) = Y(t)-
b(\bar{Z}(t-1),\bar{X}(t-1))$ where
\begin{eqnarray*}
&& b\bigl(\bar{Z}(t-1),\bar{X}(t-1)\bigr)
\\
&&\quad = \sum_{k=0}^{t-1}c\bigl(t,k,\bar
{Z}(t-1),\bar{X}(k)\bigr)
\\
&&\hspace*{35pt} {}\cdot f\bigl[1-Z(k)|\bar{Z}(k-1),\bar{X}(k)\bigr]
\end{eqnarray*}
and $f[z(k)|\bar{z}(k-1),\bar{x}(k)] = \hatPr[Z(k)=z(k)|
\bar{Z}(k-1)=\bar{z}(k-1),\bar{X}(k)=\bar{x}(k)]$ is an estimate of
the conditional probability of the observed treatment
based on some fitted parametric model (\cite{Brumback2004}). Provided
this parametric model and $c$ are both correctly specified, this
bias adjusted estimator, say $\tilde{\eta}_1$, is consistent for
$\beta_1$. Sensitivity analysis proceeds by examining how
$\tilde{\eta}_1$ changes when varying sensitivity parameters in
$c(t,k,\bar z (t-1) ,\bar x (k))$.

Because $c(t,k,\bar z (t-1) ,\bar x (k))$ is not identifiable from the
observable data, \citet{Robins1999}
recommends choosing a particular $c$ that is easily
explainable to subject matter experts to facilitate eliciting
plausible ranges of the sensitivity parameters. As an example of a
particular $c$,
\citet{Brumback2004} suggest $c(t,k,\bar{z}(t-1),\bar{x}(k)) =
\gamma\{2z(k)-1\}$ where $\gamma$ is an unidentifiable sensitivity analysis
parameter. Note that $c(t,k,\bar{z}(t-1),\bar{x}(k)) = \gamma$
for $z(k)=1$ and $c(t,k
,\bar{z}(t-1),\bar{x}(k)) = -\gamma$ for $z(k)=0$. Thus, $\gamma
>0$ ($\gamma<0$)
corresponds to subjects receiving treatment at time $k$ having greater
(smaller) mean potential outcomes at future visit $t$ than those who
did not
receive treatment at visit $k$. When $\gamma=0$, $Y(t)=Y^\gamma(t)$
and therefore $\tilde\eta_1 = \hat\eta_1$. The function
$c$ might depend on the baseline covariates $X(0)$ or the time-varying
covariates $\bar{X}(k)$. In this case, as in Section~\ref{sec:covadjust}, care should be taken in clearly
communicating the sensitivity parameters' relationship to these
covariates when eliciting plausible ranges from subject matter experts.
Another consideration when choosing a function $c$ is
whether it will allow for the sharp null of
no treatment effect, that is, for all individuals
$Y(\bar{z},t)=Y(\bar{z}',t)$ for all $\bar{z},\bar{z}'$, $t$. The
example function $c$ presented above allows for the sharp null. See
\citet{Brumback2004} for other example $c$ functions and further discussion
of sensitivity analysis for marginal structural models.

%%%%%%%%%%%%%%%%%%%%%%%%%%%%%%%%%%%%%%%%%%%%%%%%%%%%%%%%%%%%%%%%%%%%%%%%%%%%%%%%%%%%%%%%%%%%%%%%%%%%%%%%%%%%%%%%%%%
%%%%%%%%%%%%%%%%%%%%%%%%%%%%%%%%%%%%%%%%%%%%%%%%%%%%%%%%%%%%%%%%%%%%%%%%%%%%%%%%%%%%%%%%%%%%%%%%%%%%%%%%%%%%%%%%%%%
%%%%%%%%%%%%%%%%%%%%%%%%%%%%%%%%%%%%%%%%%%%%%%%%%%%%%%%%%%%%%%%%%%%%%%%%%%%%%%%%%%%%%%%%%%%%%%%%%%%%%%%%%%%%%%%%%%%
%%%%%%%%%%%%%%%%%%%%%%%%%%%%%%%%%%%%%%%%%%%%%%%%%%%%%%%%%%%%%%%%%%%%%%%%%%%%%%%%%%%%%%%%%%%%%%%%%%%%%%%%%%%%%%%%%%%
%%%%%%%%%%%%%%%%%%%%%%%%%%%%%%%%%%%%%%%%%%%%%%%%%%%%%%%%%%%%%%%%%%%%%%%%%%%%%%%%%%%%%%%%%%%%%%%%%%%%%%%%%%%%%%%%%%%
%%%%%%%%%%%%%%%%%%%%%%%%%%%%%%%%%%%%%%%%%%%%%%%%%%%%%%%%%%%%%%%%%%%%%%%%%%%%%%%%%%%%%%%%%%%%%%%%%%%%%%%%%%%%%%%%%%%

%s7 #&#
\section{Ignorance and Uncertainty Regions}\label{sec:irur}

Treatment effect bounds describe ignorance due to partial
identifiability but do
not account for uncertainty due to sampling error. This section
discusses some
methods to appropriately quantify uncertainty due to sampling
variability when
drawing inference about partially identifiable treatment effects. Over
the past
decade, a growing body of research, especially in econometrics, has considered
inference of partially identifiable parameters. The approach presented below
draws largely upon \citet{Vansteelandt2006}, who considered methods for
quantifying uncertainty in the general setting where missing data
causes partial
identifiability. As questions about treatment (or causal) effects can
be viewed
as missing data problems, the approach of Vansteelandt et al. generally applies
(under certain assumptions) to the type of problems considered
throughout this
paper. This approach builds on earlier work by \citet{Robins1997} and others.

%s7.1 #&#
\subsection{Ignorance Regions}\label{sec:ir}

Let $\boldL$
be a vector containing the potential outcomes for an individual,
let $\boldO$ denote the observed data vector, and let $\boldR$ be a vector
containing indicator variables denoting whether the corresponding
component of
$\boldL$ is observed. For example, $\boldL= (Y(1), Y(0))$, $\boldO=
(Z, Y)$,
and $\boldR= (Z, (1-Z))$ for the scenario described in Section~\ref{sec:unmeasconfound} and $\boldL = (Y(1),
Y(0), S(1), S(0))$, $\boldO = (Z,Y,S)$ and $\boldR =
(Z,(1-Z),Z,(1-Z))$ for the scenario described
in Section~\ref{sec:princstrat}. Denote the
distribution of $(\boldL,\boldR)$ by $f(\boldL,\boldR)$ and let
$f(\boldL) =
\int f(\boldL,\boldR)\,d\boldR$. The goal is to draw inference about
a parameter
vector $\bbeta$ which is a functional of the distribution of potential
outcomes $\boldL$; this is sometimes made explicit by writing $\bbeta=
\bbeta\{f(\boldL)\}$. Denote the true distribution of $(\boldL
,\boldR)$ by
$f_0(\boldL,\boldR)$ and the true value of $\bbeta$ by $\bbeta_0 =
\bbeta\{f_0(\boldL)\}$. For example, $\beta_0 = E[Y(1)-Y(0)]$ for the
scenario described in Section~\ref{sec:unmeasconfound} and $\beta
_0=E[Y(1)-Y(0)|S^{P_{0}}=(1,1)]$ for the
scenario described in Section~\ref{sec:princstrat}. Denote the true
observed data distribution by
$f_0(\boldO) = \int f_0(\boldL,\boldR)\,d\boldL_{(1-\boldR)}$ where
$\boldL_{(1-r)}$ denotes the missing part of $\boldL$ when $\boldR$ =
$r$ (i.e., the unobserved potential outcomes). The challenge in drawing
inference about
$\bbeta_0$ is that there may be multiple full data distributions
$f(\boldL,\boldR)$ that marginalize to the
true observed data distribution, that is, $
f_0(\boldO) =
\int f(\boldL,\boldR)\,d\boldL_{(1-\boldR)}
$ for some $f \neq f_0$. When this occurs, $\bbeta$ may be only partially
identifiable from $\boldO$, in which case bounds can be derived for
$\bbeta_0$ as illustrated in the sections above.

The set of values of $\bbeta\{f(\boldL)\}$ such that $f(\boldL
,\boldR)$ marginalizes to the true observed
data distribution is sometimes called the ignorance region or the
identified set.
These ignorance regions or intervals
are distinct from traditional confidence intervals in that  as the sample
size tends to infinity these intervals will not shrink to a single
point when
$\bbeta$ is partially identifiable. The ignorance region for $\bbeta$ can
be defined formally as follows. Following \citet{Robins1997},
define a class $\mathcal{M}(\gamma)$ of full data laws indexed by some
sensitivity parameter vector $\gamma$ to be nonparametrically identified
if for each observed data law $f(\boldO)$ there exists a unique law
$f(\boldL,
\boldR; \gamma)\in\mathcal{M}(\gamma)$ such that $f(\boldO)=\int
f(\boldL,\boldR;\gamma)\,d\boldL_{(1-\boldR)}$. In other words, the class
$\mathcal{M}(\gamma)$ contains a unique distribution that
marginalizes to
each possible observed data distribution. For example, for the
sensitivity analysis approach in Section~\ref{sec:pesensanal},
\citeauthor{Hudgens2006} (\citeyear{Hudgens2006}, \S4.3.3)
defined a class of full data laws
indexed by $\gamma$ given in (\ref{eq:selectionmod1})
that is nonparametrically identified. The ignorance region for $\bbeta
$ is formally defined to be
%
%e37 #&#
\begin{eqnarray}
\label{eq:ir} &&\ir_{f_{0}}(\bbeta,\bGamma)
\nonumber
\\
&&\quad = \biggl\{\vphantom{\int} \bbeta\bigl\{f(\boldL)\bigr\} : f(\boldL)
\nonumber
\\
&&\hspace*{18pt}\quad = \int{f(\boldL,\boldR;\gamma)\,d\boldR} \mbox{ for some }
\\
&&\hspace*{27pt} {} f(\boldL,\boldR)\in\mathcal{M}(\bGamma) \mbox{ such that }
\nonumber
\\
&&\hspace*{27pt} {} \int{f(\boldL},\boldR;\gamma)\,d\boldL_{(1-\boldR)}
= f_0(\boldO) \biggr\},
\nonumber
\end{eqnarray}
where $\bGamma$ is the set of all possible values of $\gamma$ under
whatever set of assumptions is being invoked and $\mathcal{M}(\bGamma
) = \bigcup_{\gamma\in\bGamma}
\mathcal{M}(\gamma)$. Assume $\mathcal{M}(\bGamma)$ contains the
true full
data distribution, that is, $f_0(L,R)=f(L,R,\gamma_0)$ for some $\gamma
_0\in\bGamma$.
[For considerations when $\mathcal{M}(\Gamma)$ does not contain the
true full data distribution, see \citet{Todem2010}.]
Because $\mathcal{M}(\gamma)$ is nonparametrically identified, for
each $\gamma\in\bGamma$ there is a
single $\bbeta(\gamma) =  \bbeta\{\int f(\boldL,\boldR;
\gamma)\,d\boldR)\}$ in the ignorance region (\ref{eq:ir}). If
$\mathcal{M}(\Gamma)$
includes all possible full data distributions that marginalize to any
possible observed
data distribution, then the ignorance region will contain the bounds.

In practice, the ignorance region will be unknown because it depends on
the unknown true observed data
distribution $f_0(\boldO)$.
For $\gamma$ fixed, $\bbeta(\gamma)$ is identifiable from the
observed data and
the ignorance region can be estimated by estimating $\bbeta(\gamma)$ for
each value of $\gamma\in\bGamma$, denoted by
$\hat{\bbeta}(\gamma)$.
The resulting estimator of $\ir_{f_{0}}(\bbeta,
\bGamma)$ is then $\{\hat{\bbeta}(\gamma)\dvtx  \gamma\in
\bGamma\}$. For scalar $\beta(\gamma)$, let $\hat{{\beta
}}_{l}=\inf_{\gamma\in
\bGamma}\{\hat{{\beta}}(\gamma)\}$ and $\hat{{\beta}}_{u}=
\sup_{\gamma\in\bGamma}\{\hat{{\beta}}(\gamma)\}$ such that the
estimated ignorance region is contained in the interval $[\hat\beta
_l, \hat
\beta_u]$.

%s7.2 #&#
\subsection{Uncertainty Regions}\label{sec:ur}

Estimated ignorance regions convey ignorance due to partial
identifiability and
do not reflect sampling variability in the estimates. Indeed much of the
literature on bounds and sensitivity analysis of treatment effects
tends to
report estimated ignorance regions and either ignores sampling
variability or
employs ad-hoc inferential approaches such as pointwise confidence intervals
conditional on each value of the unidentifiable sensitivity parameter. More
recent developments have provided a formal framework for conducting
inference in
partial identifiability settings
(\cite{Imbens2004}; \cite{Vansteelandt2006}; \cite{Romano2008}; \cite{Bugni2010}; \cite{Todem2010}).
The main focus in this research has been the construction of confidence
regions for either
the parameter $\bbeta_0$ or the ignorance region $\ir_{f_{0}}(\bbeta_0,
\bGamma)$.

Following \citet{Vansteelandt2006}, a $(1-\alpha)$ pointwise uncertainty
region for $\bbeta_0$ is defined to be a region $\UR_p(\bbeta,
\bGamma)$ such that
\[
\inf_{\gamma\in\bGamma}\Prr_{f_{0}} \bigl\{\bbeta(\gamma) \in
\UR_p(\bbeta,\bGamma) \bigr\} \geq1 - \alpha,
\]
where $\Prr_{f_{0}} \{\cdot \}$ denotes probability
under $f_{0}(\boldO)$. That is, $\UR_p(\bbeta,\bGamma)$ contains
$\bbeta(\gamma)$ with at least probability $1-\alpha$ for all
$\gamma\in\bGamma$. In particular, assuming $\gamma_0\in\bGamma$,
then $\UR_p(\bbeta,\bGamma)$ will contain $\bbeta_0=\bbeta(\gamma
_0)$ with at
least probability $1-\alpha$.

An appealing aspect of pointwise uncertainty regions is that they
retain the
usual duality between confidence intervals and hypothesis testing.
Namely, one
can test the null hypothesis $H_0\dvtx  \bbeta_0 = \bbeta_c$ versus $H_a\dvtx
\bbeta_0
\neq\bbeta_c$ for some specific $\bbeta_c$ at the $\alpha$
significance level
by rejecting $H_0$ when the $(1-\alpha)$ pointwise uncertainty region
$\UR_p(\bbeta,\bGamma)$ excludes $\bbeta_c$. This is easily shown by
noting for
$\bbeta_c=\bbeta(\gamma_0)$
\begin{eqnarray*}
&&\Prr_{f_{0}}[\mbox{reject }H_0]
\\
&&\quad = 1-\Prr_{f_{0}} \bigl\{\bbeta(\gamma_0) \in
\UR_p(\bbeta,\bGamma) \bigr\}
\\
&&\quad \leq1-{\inf}_{\gamma\in\bGamma} \Prr_{f_{0}} \bigl\{ \bbeta(\gamma)
\in\UR_p(\bbeta,\bGamma) \bigr\} \leq\alpha,
\end{eqnarray*}
where the last inequality follows because $\UR_p(\bbeta,\bGamma)$ is a
$(1-\alpha)$ pointwise uncertainty region.

Various methods under different assumptions have been proposed for constructing
pointwise uncertainty regions. \citet{Imbens2004} and \citet{Vansteelandt2006}
proposed a simple method for constructing pointwise uncertainty
regions for a scalar $\beta$ with ignorance region $[\beta_l, \beta_u]$.
Let $\gamma_l, \gamma_u \in\bGamma$ be the values of the sensitivity
parameter such that $\beta_l=\beta(\gamma_l)$ and $\beta_u=\beta
(\gamma_u)$.
Assume
%
%e38 #&#
%e39 #&#
\begin{eqnarray}
\label{eq:assume1} %
&&\mbox{There exist } \hat\beta_l \mbox{ such that }
\nonumber
\\[-2pt]
&&\sqrt{n}(\hat\beta_l - \beta_l)
\stackrel{d}{\to}  N
\bigl(0,\sigma^2_l\bigr)
\nonumber
\\[-10pt]
\\[-8pt]
&&\mbox{ and } \hat\beta_u \mbox{ such that }
\nonumber
\\[-2pt]
&&\sqrt{n}(\hat\beta_u - \beta_u) \stackrel{d}{\to} N
\bigl(0,\sigma^2_u\bigr).
\nonumber
\label{eq:assume2}
\\
&&\mbox{The values }\gamma_l \mbox{ and } \gamma_u
\mbox{ are the same}
\nonumber
\\[-13pt]
\\[-8pt]
&& \mbox{for all possible observed data laws.}
\nonumber
\end{eqnarray}
Under assumptions (\ref{eq:assume1}) and (\ref{eq:assume2}),
an asymptotic $(1-\alpha)$ pointwise uncertainty interval for $\beta
_0$ is
%
%e40 #&#
\begin{eqnarray}
\label{eq:ur2} &&\UR_p(\beta,\bGamma)
\nonumber
\\[-8pt]
\\[-8pt]
&&\quad = [\hat{\beta_l}-c_\alpha\widehat{
\sigma}_l/\sqrt{n}, \hat{\beta}_u+c_\alpha
\widehat{\sigma}_u/\sqrt{n} ],
\nonumber
\end{eqnarray}
where $c_\alpha$ satisfies
%
%e41 #&#
\begin{equation}
\label{eq:ur4} \Phi \biggl(c_\alpha+\frac{\sqrt n (\hat{\beta}_u-\hat{\beta
_l} )}{\max
\{\widehat{\sigma}_l,\widehat{\sigma}_u\}} \biggr)-
\Phi(-c_\alpha)=1-\alpha,
\end{equation}
$\Phi(\cdot)$ denotes the cumulative distribution function of a
standard normal
variate, and $\widehat{\sigma}_l$ and $\widehat{\sigma}_u$ are
consistent estimators of
$\sigma_l$ and $\sigma_u$, respectively
(\cite{Imbens2004}; \cite{Vansteelandt2006}). Note if $\hat{\beta}_u - \hat
{\beta}_l>0$ and $n$ is large such that the left-hand side of
(\ref{eq:ur4}) is approximately equal to $1-\Phi(-c_\alpha)$, then
$c_\alpha
\approx z_{1-\alpha}$, the $(1-\alpha)$ quantile of a standard normal
distribution. In contrast, if $\hat{\beta}_u=\hat{\beta}_l$, then
$c_\alpha = z_{1-\alpha/2}$.

In addition to the pointwise uncertainty region, \citet{Horowitz2000} and
\citet{Vansteelandt2006} define a $(1-\alpha)$ strong uncertainty
region for
$\bbeta_0$ to be a
region $\UR_s(\bbeta,\bGamma)$ such that
\[
\Prr_{f_{0}} \bigl\{ \ir_{f_{0}}(\bbeta,\bGamma) \subseteq
\UR_s(\bbeta,\bGamma) \bigr\} \geq1 - \alpha,
\]
that is, $\UR_s(\bbeta,\bGamma)$ contains the entire ignorance region
with probability at least $1-\alpha$. Whereas the pointwise
uncertainty region
can be viewed as a confidence region for the partially identifiable target
parameter $\bbeta_0$, the strong uncertainty region is a confidence
region for
the ignorance region $\ir_{f_{0}}(\bbeta,\bGamma)$. Clearly, any
strong uncertainty
region will also be a (conservative) pointwise uncertainty region as
$\bbeta_0
\in\ir_{f_{0}}(\bbeta,\bGamma)$. Under assumptions (\ref
{eq:assume1}) and
(\ref{eq:assume2}), an asymptotic $(1-\alpha)$ strong uncertainty
interval for
scalar $\beta_0$ is simply
%
%e42 #&#
\begin{eqnarray}
\label{eq:ur2strong} &&\UR_s(\beta,\bGamma)
\nonumber
\\[-8pt]
\\[-8pt]
&&\quad = [\hat{\beta_l}-z_{1-{\alpha}/{2}}\widehat{
\sigma}_l/\sqrt{n}, \hat{\beta}_u+z_{1-{\alpha}/{2}}
\widehat{\sigma}_u/\sqrt{n} ].
\nonumber
\end{eqnarray}
Note that (\ref{eq:ur2strong}) is equivalent to the union of all
pointwise $(1-\alpha)$ confidence
intervals for $\bbeta(\gamma)$ under $\mathcal{M}(\gamma)$ over all
$\gamma
\in\bGamma$, which is a simple approach often employed when reporting
sensitivity
analysis. Because strong uncertainty intervals are necessarily
pointwise intervals, this
simple approach is also a valid method for computing pointwise
intervals, although intervals based on
(\ref{eq:ur2}) will always be as or more narrow.

The two key assumptions (\ref{eq:assume1}) and (\ref{eq:assume2}) may
not hold in general. For example,
(\ref{eq:assume1}) may not hold for all possible observed data
distributions, particularly for extreme values of $\gamma_l$
or $\gamma_u$. Assumption (\ref{eq:assume2}) may not hold if
different observed data distributions place different
constraints on the possible range of $\gamma$ or if $\Gamma$ is
chosen by the data analyst on the basis of the observed data.
If (\ref{eq:assume1}) or (\ref{eq:assume2}) does not hold,
alternative inferential
methods are needed
(e.g., see \cite{Vansteelandt2001}; \cite{Horowitz2006}; \cite{Chernozhukov2007}; \cite{Romano2008}; \cite{Stoye2009}; \cite{Todem2010}; \cite{Bugni2010}).

A third approach to quantifying uncertainty due to sampling variability
is to
consider $\bbeta(\cdot)$ as function of $\gamma$ and construct a
$(1-\alpha)$
simultaneous confidence band for the function $\bbeta(\cdot)$. That
is, a random
function $\CB(\cdot)$ is found such that
\[
\Prr_{f_{0}} \bigl\{ \bbeta(\gamma) \in\CB(\gamma) \mbox{ for all }
\gamma\in \bGamma \bigr\} \geq1 - \alpha.
\]
It follows immediately that $\bigcup_{\gamma\in\bGamma} \CB(\gamma)$ is a
strong uncertainty region (and thus a pointwise uncertainty region as well).
\citet{Todem2010} suggest a bootstrap approach to constructing confidence
bands.

Whether pointwise uncertainty regions, strong uncertainty regions, or
confidence bands are preferred will be context specific. Typically, it
is of interest to
draw inference about a single target parameter and not the entire
ignorance region.
Thus, in general pointwise uncertainty regions may have greater utility
than strong
uncertainty regions.
Because strong uncertainty regions are necessarily conservative
pointwise uncertainty
regions, the strong regions can be useful in settings where determining
a pointwise
region is more difficult. Additionally, in
some settings it may be of interest to assess whether $\bbeta$ is
nonzero, for example,
if $\bbeta$ denotes the effect of treatment. In these settings,
computing a
confidence band $\CB(\cdot)$ has the advantage of providing the
subset of
$\bGamma$ where the null hypothesis $\bbeta(\gamma)=0$ can be
rejected. This
is especially appealing if $\gamma$ is scalar, in which case a
confidence band
(as in Figure~3 of  \cite{Todem2010}) provides a simple
approach to reporting sensitivity analysis results. On the other hand, if
$\gamma$ is multidimensional, visualizing confidence bands can be difficult
and instead reporting the (pointwise or strong) uncertainty region may
be more practical.

%s7.3 #&#
\subsection{Data Example}\label{sec:revisit}

%f1 #&#
\begin{figure}

\includegraphics{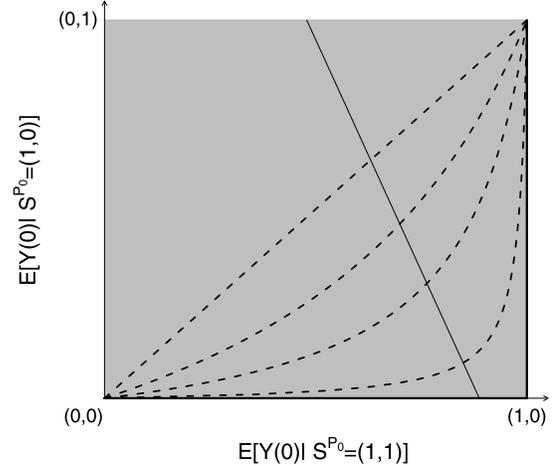}

\caption{Graphical depiction of the bounds and sensitivity analysis model described in Sections \protect\ref{sec:pebounds}
and \protect\ref{sec:pesensanal}.
The solid thin line with negative slope represents
a set of joint distribution functions of $(Z,S(1),S(0),Y(1),Y(0))$
that all
give rise to the same distribution of the observable random variables $(Z,S,Y)$.
The four dotted curves depict the log odds ratio selection model for $\gamma=0,1,2,4$.
The $\gamma=0$ model is equivalent to the no selection model.
Each selection model identifies exactly one pair of expectations
from this set, rendering the principal effect (\protect\ref{eq:ACEy1}) identifiable. The thick black lines on the edge
of the unit square correspond to the lower bound of the principal effect.}\label{fig:statsci1}
\end{figure}

\begin{figure}[t]

\includegraphics{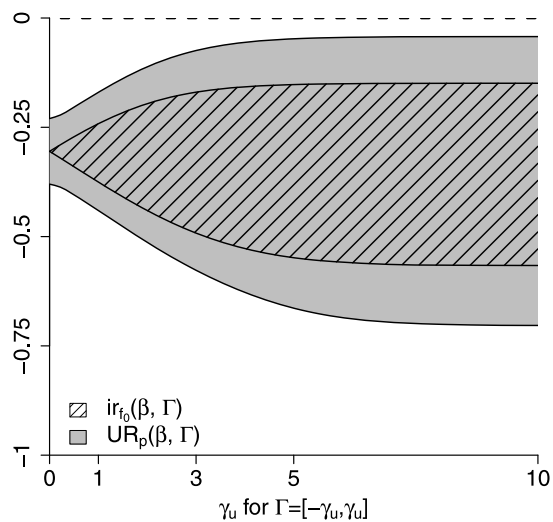}

\caption{Estimated ignorance regions
$\ir_{f_{0}}(\beta,\Gamma)$ and 95\% pointwise uncertainty regions
$\UR_p(\beta,\Gamma)$ for the pertussis vaccine example in Section \protect\ref{sec:revisit}. The
principal effect (\protect\ref{eq:ACEy1}) is denoted $\beta$ and $\bGamma=[-\gamma_u,\gamma_u]$ for $\gamma_u$ along
the horizontal axis. The curve given by the lower boundary of the area
with black slanted lines corresponds to $\hat{\beta}_l$, the minimum of the estimated ignorance regions, and
the upper bound of the area with black slanted lines corresponds to $\hat{\beta}_u$, the maximum
of the estimated ignorance region. The curve given by the lower (upper) boundary of the
gray shaded area corresponds to the minimum (maximum) of the 95\% pointwise uncertainty region.}\label{fig:statsci2}
\end{figure}

Returning to the pertussis vaccine study described in Section~\ref{sec:princstrat}, an analysis that ignores the
potential for selection bias might entail computing a naive estimator
(the difference in
empirical means of $Y$ between the vaccinated and unvaccinated amongst
those infected) along with a
95\% Wald confidence interval, which would be $-$0.31 (95\% CI $-$0.38, $-$0.23).
If the sensitivity analysis approach in Section~\ref{sec:pesensanal}
is applied, the parameter of
interest $\beta(\gamma)= E[Y(1)-Y(0)|S^{P_0}=(1,1)]$ is identified
for fixed
values of the sensitivity analysis parameter $\gamma$ given in
(\ref{eq:selectionmod1}). For fixed $\gamma$, $E[Y(0)|S^{P_0}=(1,1)]$
is determined by the intersection of the
negative sloped line (\ref{eq:id1}) and the curve (\ref
{eq:selectionmod1}), which is illustrated in Figure~\ref{fig:statsci1} for
the pertussis data.
Because $E[Y(0)|S^{P_0}=(1,1)]$ increases with $\gamma$,
$\beta(\gamma)$ is a monotonically decreasing function of $\gamma$.
Therefore $\gamma_l$ and
$\gamma_u$ equal the maximum and minimum values of $\bGamma$
regardless of the
observed data law, indicating (\ref{eq:assume2}) holds provided that
$\Gamma$ is chosen by the analyst independent of the observed data.
For $\gamma$ fixed and finite, $\beta(\gamma)$ can be estimated via
nonparametric maximum likelihood (i.e., without any additional assumptions).
This estimator will be consistent and asymptotically
normal under standard regularity conditions if $\Prr[S(0)>S(1)]>0$ (i.e.,
the vaccine has a
protective effect against infection). For $\gamma=\pm\infty$ and
$\Prr[S(0)>S(1)]>0$, \citet{Lee2009} proved that the
estimators of the bounds similar to those given in Section~\ref{sec:pebounds} are consistent and
asymptotically normal for a continuous outcome $Y$. The limiting
distribution of the estimator of the upper bound ($\gamma=-\infty$)
for a binary outcome will be normal if in addition
%
%e43 #&#
\begin{equation}
\label{eq:asynormfail1} 1-E[Y|S=1, Z=0] \neq\frac{\Prr[S=1|Z=1]}{\Prr[S=1|Z=0]},
\end{equation}
and similarly the estimator of the lower bound ($\gamma=\infty$) will
be asymptotically normal if in addition
%
%e44 #&#
\begin{equation}
\label{eq:asynormfail2} E[Y|S=1, Z=0] \neq\frac{\Prr[S=1|Z=1]}{\Prr[S=1|Z=0]}.
\end{equation}
Likelihood ratio tests for the null hypotheses that (\ref
{eq:asynormfail1}) and (\ref{eq:asynormfail2})
do not hold yield p-values $p<10^{-4}$ and $p=0.18$, respectively,
indicating strong evidence that (\ref{eq:asynormfail1})
holds and equivocal evidence regarding (\ref{eq:asynormfail2}).
Assuming (\ref{eq:asynormfail1}) and (\ref{eq:asynormfail2})
both hold implies (\ref{eq:assume1}), such that (\ref{eq:ur2})
and (\ref{eq:ur2strong}) can be used to construct $(1-\alpha)$
pointwise and
strong uncertainty intervals for $\beta_0$. Estimated ignorance and uncertainty
intervals of $\beta_0$ for different choices of $\bGamma$ are given
in Table~\ref{tab:statsci1}
and Figure~\ref{fig:statsci2}, with standard error estimates obtained
using the observed
information.
Even for $\bGamma= (-\infty,\infty)$ both the pointwise and strong
uncertainty
intervals exclude zero, indicating a significant effect of vaccination. In
particular, with 95\% confidence we can conclude the vaccine decreased
the risk
of severe disease among individuals who would have become infected
regardless of
vaccination. \nocite{Lee2009}

%t1 #&#
\begin{table}[t]
\tabcolsep=0pt
\caption{Pertussis vaccine study data: Estimated ignorance regions
and 95\% pointwise and strong uncertainty regions
of $\beta= E[Y(1)-Y(0)|S^{P_0}=(1,1)]$ for different $\bGamma$}
\label{tab:statsci1}
\begin{tabular*}{\columnwidth}{@{\extracolsep{\fill}}lccc@{}}
\bhline
${\bGamma}$          & ${\ir_{f_{0}}(\beta,\bGamma)}$ & ${\UR_p(\beta,\bGamma)}$ & ${\UR_s(\beta,\bGamma)}$ \\
\hline
$[-3,3]$           & $[ - 0.49,  - 0.17]$               & $[ - 0.58,  - 0.07]$         & $[ - 0.59,  - 0.06]$         \\
$[-5,5]$          & $[ - 0.55,  - 0.15]$               & $[ - 0.66,  - 0.05]$         & $[ - 0.69,  - 0.03]$         \\
$[-10,10]$        & $[ - 0.57,  - 0.15]$               & $[ - 0.70,  - 0.04]$         & $[ - 0.73,  - 0.02]$         \\
$(-\infty,\infty)$ & $[ - 0.57,  - 0.15]$               & $[ - 0.70,  - 0.04]$         & $[ - 0.73,  - 0.02]$         \\
\hline
\end{tabular*}
\end{table}

%%%%%%%%%%%%%%%%%%%%%%%%%%%%%%%%%%%%%%%%%%%%%%%%%%%%%%%%%%%%%%%%%%%%%%%%%%%%%%%%%%%%%%%%%%%%%%%%%%%%%%%%%%%%%%%%%%%
%%%%%%%%%%%%%%%%%%%%%%%%%%%%%%%%%%%%%%%%%%%%%%%%%%%%%%%%%%%%%%%%%%%%%%%%%%%%%%%%%%%%%%%%%%%%%%%%%%%%%%%%%%%%%%%%%%%
%%%%%%%%%%%%%%%%%%%%%%%%%%%%%%%%%%%%%%%%%%%%%%%%%%%%%%%%%%%%%%%%%%%%%%%%%%%%%%%%%%%%%%%%%%%%%%%%%%%%%%%%%%%%%%%%%%%
%%%%%%%%%%%%%%%%%%%%%%%%%%%%%%%%%%%%%%%%%%%%%%%%%%%%%%%%%%%%%%%%%%%%%%%%%%%%%%%%%%%%%%%%%%%%%%%%%%%%%%%%%%%%%%%%%%%
%%%%%%%%%%%%%%%%%%%%%%%%%%%%%%%%%%%%%%%%%%%%%%%%%%%%%%%%%%%%%%%%%%%%%%%%%%%%%%%%%%%%%%%%%%%%%%%%%%%%%%%%%%%%%%%%%%%
%%%%%%%%%%%%%%%%%%%%%%%%%%%%%%%%%%%%%%%%%%%%%%%%%%%%%%%%%%%%%%%%%%%%%%%%%%%%%%%%%%%%%%%%%%%%%%%%%%%%%%%%%%%%%%%%%%%

%f2 #&#

%s8 #&#
\section{Discussion}\label{sec:statscidiscuss}

This paper considers conducting inference about the effect of a
treatment (or
exposure) on an outcome of interest. Unless treatment is randomly
assigned and
there is perfect compliance, the effect of treatment may be only partially
identifiable from the observable data. Through the five settings in
Sections~\ref{sec:unmeasconfound}--\ref{sec:longtrt}, we discussed
two approaches often employed to address partial
identifiability: (i) bounding the treatment effect under minimal
assumptions, or (ii)
invoking additional untestable assumptions that render the treatment
effect identifiable and
then conducting sensitivity analysis to assess how inference about the treatment
effect changes as the untestable assumptions are varied. Incorporating
uncertainty due to sampling variability was discussed in Section~\ref{sec:irur}, and
throughout large-sample frequentist methods were considered. Analogous Bayesian
approaches to partial identification
(\cite{Gustafson2010}; \cite{Moon2011}; \cite{Richardson2011})
and sensitivity analysis (\cite{McCandless2007}; \cite{Gustafson2010a}) have
also been developed.

Determining treatment effect bounds is essentially a constrained
optimization problem, where the
constraints are determined by the relationship between the
distributions of the observable
random variables and of the potential outcomes under whichever
assumptions are being made.
In simple cases, such as in Section~\ref{sec:noassbounds}, bounds can easily
be derived from first principles and may have simple closed forms; in
more complicated settings,
such as in Section~\ref{sec:compliance}, bounds may be determined
using linear programming or other optimization
methods.
In many cases, calculating bounds under minimal assumptions may seem to
be a
meaningless exercise because the bounds are often quite wide and may
not exclude
the null of no treatment effect as seen with the ``no assumptions''
bounds in
Section~\ref{sec:unmeasconfound}. On the contrary, in settings like this
\citet{Robins1996} write: ``Some argue against reporting bounds for
nonidentifiable parameters, because bounds are often so wide as to be useless
for making public health decisions. But we view the latter problem as a reason
\textit{for} reporting bounds in conjunction with other analyses: Wide
bounds make
clear that the degree to which public health decisions are dependent on merging
the data with strong prior beliefs.''

Bounds may be narrowed by reducing the feasible region of the
optimization problem.
This may be accomplished by considering further assumptions that place
restrictions on either
the distributions of the potential outcomes, the distributions of the
observable random variables,
or both. Assumptions that place restrictions on the observable random variables
may have implications which are testable. If the observed data provide
evidence against any
assumptions being considered, bounds should be computed without making
these assumptions. Those assumptions
without testable implications can only be determined to be plausible or
not by subject matter experts.

A potentially less conservative approach to computing bounds is to make
untestable assumptions which identify the causal estimand and then
assess the
robustness of inference drawn to departures from these assumptions in a
sensitivity analysis.
A general guideline for
specifying the sensitivity analysis parameters representing these
departures is to choose parameters that are easily
interpretable to subject matter experts. Parameter specification will
depend on whether or not sensitivity analysis is conducted
by directly modeling the association of an unmeasured confounder $U$
with treatment selection and the potential outcomes.
Sensitivity analyses based on this approach are applicable when the
existence of $U$ is known and there is some historical knowledge of the
magnitude
association of $U$ with $Z$ and the potential outcomes
(\cite{Robins1999}; \cite{Brumback2004}).
Otherwise, alternative approaches based on directly modeling the
unobserved potential outcome distributions may be preferred.
A second guiding principle should be to avoid
specifications of sensitivity parameters that place restrictions on the
distributions of
observable random variables that are not empirically supported. A third
consideration when conducting sensitivity analysis concerns determining a
plausible region of the sensitivity parameters. That the region be
chosen prior to
data analysis is in general necessary for inference, such as described
in Section~\ref{sec:irur}, to be valid. Choice of the region of the sensitivity
parameters may be
dictated by whether one wants to consider only mild or also severe
departures from the
identifying assumptions. If the
identifying assumption in question is considered plausible, then it may
be that
only mild departures from the assumption are deemed necessary for the
sensitivity
analysis. In this case, subject matter experts can be consulted to determine,
prior to data analysis, a plausible region for the sensitivity
parameters. If, on the other hand, severe
departures from untestable identifying assumptions are to be
entertained, sensitivity analyses
should be conducted over all possible values of the sensitivity
parameters. Sensitivity analyses which consider all possible
full data distributions that marginalize to the observed data
distribution will yield ignorance regions containing the bounds.

Though the examples presented here demonstrate the broad scope of
scenarios where
bounds and sensitivity analysis methods have been derived and employed
to draw inference
about treatment effects, they
certainly are not exhaustive of all settings where these methods have been
developed. For instance, \citet{VanderWeele2011c} consider
sensitivity analysis to unmeasured confounding for causal interaction effects.
Bounds and sensitivity analysis methods have also
recently been considered in the presence of interference, that is, in
settings where
treatment of one individual may affect the outcome of another
individual, such as
in social networks (\cite{Steeg2010}; \cite{VanderWeele2011b}; \cite{Manski2013}). For
studies where
sensitivity analyses are planned or anticipated, Rosenbaum and
colleagues have
examined how aspects of study design and the choice of statistical
tests or
estimators may affect the power or precision of the sensitivity
analyses to be
conducted (\cite{Heller2009}; \citeauthor{Rosenbaum2010a}, \citeyear{Rosenbaum2010a}; \citeyear{Rosenbaum2010b}; \citeyear{Rosenbaum2011}).

Bounds and sensitivity analyses of treatment effects have been utilized
in various substantive
settings, such as biomedical research
(e.g., \cite{Cole2005}; \cite{Rerks2009}; \cite{VanderWeele2011d}; \cite{Hu2012})
and economics (e.g., \cite{Heckman2001}; \cite{Sianesi2004}; \cite{Armstrong2010}).
Nonetheless, despite the wide range of settings in which these methods
are applicable, their use in substantive settings remains somewhat
limited in frequency.
Given the large amount of literature detailing their broad scope of
applicability and
that formal inferential methods for partially identifiable parameters
are now available,
hopefully these approaches will be employed with greater frequency in
substantive
settings in the future.

The sensitivity analyses described throughout this paper focus on
departures from untestable assumptions
which identify treatment effects. Other types of sensitivity analyses
might be considered as well, for example,
to assess how robust inferences are to various analytical decisions
that are invariably made in data analysis.
\citeauthor{Rosenbaum2002} (\citeyear{Rosenbaum2002}, Section~11.9)
 refers to such assessment as
``stability analysis,'' in contrast to the types of sensitivity analyses
discussed above. See \citeauthor{Rosenbaum1999} (\citeyear{Rosenbaum1999}, \citeyear{Rosenbaum2002}) and
\citeauthor{Morgan2007} (\citeyear{Morgan2007}, Section~6.2)
 for further discussion regarding various types
of sensitivity analyses beyond the type considered here.

\section*{Acknowledgments}
The authors were partially supported by NIH NIAID grants R01 AI085073
and R37 AI054165. The content is solely the responsibility of the
authors and
does not necessarily represent the official views of the National
Institute of
Allergy and Infectious Diseases or the National Institutes of Health. The
authors thank Guest Editors Andrea Rotnitzky and Thomas Richardson for the
invitation to contribute to this special issue of \textit{Statistical Science}; and
the Associate Editor and reviewers for their helpful comments.
 This paper was written while the first author was a Ph.D. student in the Department of Biostatistics at the University of North Carolina.

%suskaldyti doi

% imsref loaded by aiste.veprauskaite, 2014-10-16 16:18:49
% imsref loaded by aiste.veprauskaite, 2014-10-22 13:54:21
% imsref loaded by aiste.veprauskaite, 2014-10-22 13:59:41
% imsref loaded by aiste.veprauskaite, 2014-10-22 14:01:50
% imsref loaded by aiste.veprauskaite, 2014-10-22 14:02:36
% imsref loaded by aiste.veprauskaite, 2014-10-22 14:08:18
% imsref loaded by aiste.veprauskaite, 2014-10-22 14:12:25

\end{document}